\documentclass{llncs}

\usepackage{amsmath}

\usepackage{pslatex}

\usepackage{amssymb}
\usepackage{stmaryrd}
\usepackage{graphicx}
\usepackage{marvosym}
\usepackage{color}
\usepackage{pdfsync}
\usepackage{wasysym}
\usepackage{chemarr}

\usepackage{nicefrac}

\graphicspath{{figures/}} \DeclareGraphicsExtensions{.pdf,.jpg}
\usepackage[center]{subfigure}

\newcommand{\emptysys}{\mathbf 0}

\newcommand{\Real}[1]{\mathrm{Real}}

\newcommand{\coco}{\mbox{\ensuremath{\mathrm{CO}_2}\hspace{2pt}}}

\newcommand{\names}{\ensuremath{\mathcal{N}}}
\newcommand{\snames}{\names}

\newcommand{\vars}{\mathcal V}

\newcommand{\pmv}[1]{\ensuremath{\mathsf{#1}}}
\newcommand{\atom}[1]{\textup{\textsf{#1}}}

\newcommand{\fact}[2]{\mathsf{do}_{#1}\,{#2}}
\newcommand{\tell}[2]{\mathsf{tell}_{#1}\,{#2}}
\newcommand{\sys}[2]{{#1} [{#2}] }

\newcommand{\freeze}[2]{\downarrow_{#1}{#2}}

\newcommand{\sep}{\ \bnfmid\ }

\newcommand{\agreement}[4]{{#1} \vartriangleright_{#3}^{#4} {#2}}

\newcommand{\hidden}[1]{}
\newcommand{\dom}[1]{\mathrm{dom}({#1})}
\newcommand{\defeq}{\mathrel{\mathop{=}\limits^{\rm def}}}

\def\vec{\mathaccent"017E }

\newcommand{\proofend}{\mbox{$\Box$}}

\newcommand{\mmdef}{\mbox{$\;\stackrel{\textrm{\tiny def}}{=}\;$}}

\newcommand{\fn}[1]{\mathrm{fn}(#1)}

\newcommand{\fv}[1]{\mathrm{fv}(#1)}

\newcommand{\irule}[2]{
  \begin{array}{c}
    #1  \\ \hline
    #2
  \end{array}}

\newcommand{\subs}[2]{\{^{#1}/_{#2}\}}

\newcommand{\bnfmid}{\;\big|\;}

\newcommand{\imp}{\rightarrow}
\newcommand{\nrule}[1]{{\footnotesize \textsc{#1}}}

\newcommand{\ask}[2]{\mathsf{ask}_{{#1}}\,{#2}}
\newcommand{\fuse}[2][]{\mathsf{fuse}_{{#2}}}
 \newcommand{\says}{\ensuremath{\;\mathit{says}\;}}

\newcommand{\pnil}{\mathbf{0}}

\newcommand{\bind}[2]{\nicefrac{#2}{#1}}
\newcommand{\setenum}[1]{\{#1\}}
\newcommand{\setcomp}[2]{\{{#1} \,\mid\, {#2}\}}

\newcommand{\SumIntRaw}{\bigoplus}
\newcommand{\SumExtRaw}{\sum}

\newcommand{\sumInt}{\oplus}
\newcommand{\sumExt}{+}

\newcommand{\SumInt}[3][]{\SumIntRaw_{#1} {#2} \, ; \, {#3}}
\newcommand{\SumExt}[3][]{\SumExtRaw_{#1} {#2} \, . \, {#3}}

\newcommand{\sumI}[2]{{#1} \, ; \, {#2}}
\newcommand{\sumE}[2]{{#1} \, . \, {#2}}

\newcommand{\ready}[1]{\mathit{ready}\; {#1}}
\newcommand{\rec}[2]{\mathit{rec}\; {#1}.\; {#2}}
\newcommand{\cnil}{\ensuremath{0}}

\newcommand{\E}{\textit{E}}
\newcommand{\co}[1]{{\textit{co}}(#1)}
\newcommand{\rs}[1]{{\textit{RS}}(#1)}

\newcommand{\compliant}[0]{\bowtie}
\newcommand{\dual}[1]{\textit{dual}(#1)}
\newcommand{\ncompliant}[0]{\not\bowtie}

\newcommand{\cmove}[1]{\xrightarrow{#1}\hspace{-1.8ex}\rightarrow}
\newcommand{\abscmove}[1]{\cmove{#1}_{\sharp}}
\newcommand{\pmove}[2][]{\xrightarrow{#2}}
\newcommand{\abspmove}[2][]{{\xrightarrow{#2}_{\sharp}^{#1}}}

\newcommand{\csmiley}[3][]{{#2} \hspace{3pt}\dot{}\hspace{4pt}\dot{}\hspace{-6pt}{\smallsmile}\hspace{1pt}_{{#1}} {#3}}
\newcommand{\cfrown}[3][]{{#2} \hspace{3pt}\dot{}\hspace{4pt}\dot{}\hspace{-6pt}{\smallfrown}\hspace{1pt}_{{#1}} {#3}}

\newcommand{\bic}[2]{{\pmv A} \says {#1} \mid {\pmv B} \says {#2}}

\newcommand{\does}[3][]{{#2} \says \fact{#1}{#3}}

\newcommand{\unblocked}{\tau}

\newcommand{\ctx}{\mathit{ctx}}

\newcommand{\readydo}[2]{\textit{RD}_{#1}({#2})}
\newcommand{\unblocks}[2]{{#1} \;\mathit{unblocks}\; {#2}}
\newcommand{\realizes}[4][]{{#2} \models_{#4}^{#1} {#3}}
\newcommand{\notrealizes}[4][]{{#2} \not\models_{#4}^{#1} {#3}}
\newcommand{\canonical}{\text{$\sharp$-honest}}
\newcommand{\canonicity}{\text{$\sharp$-honesty}}

\newcommand{\open}[1]{open(#1)}

\newenvironment{barttheorem}[2][]{\medskip\noindent\textbf{{#2}{#1}.}\it}{}

\newtheorem{appdefinition}{Definition}[section]
\newtheorem{applemma}{Lemma}[section]

\title{On the realizability of contracts in dishonest systems}

\author{Massimo Bartoletti\inst{1} \and Emilio Tuosto\inst{2} \and Roberto Zunino\inst{3}}
\institute{Dipartimento di Matematica e Informatica, Universit\`a degli Studi di Cagliari, Italy \and Department of Computer Science, University of Leicester, UK \and DISI, Universit\`a di Trento and COSBI, Italy}

\begin{document}

\maketitle

\begin{abstract}
We develop a theory of contracting systems, where behavioural contracts 
may be violated by dishonest participants after they have been agreed upon
--- unlike in traditional approaches based on behavioural types.
We consider the contracts of~\cite{Castagna09toplas}, and we embed them in a calculus that allows distributed participants to advertise contracts, reach
agreements, query the fulfilment of contracts, and realise them (or choose not to).
Our contract theory makes explicit who is culpable 
at each step of a computation. 
A participant is honest in a given context $S$ when she is not culpable in each possible interaction with $S$.
Our main result is a sufficient criterion for classifying 
a participant as honest in all possible contexts.
\end{abstract}

\section{Introduction}

Contracts are abstract descriptions of the behaviour of services.
They are used to compose services which are \emph{compliant} according to some
semantic property, e.g.\ the absence of
deadlocks~\cite{Bravetti07sc,Carpineti06basic,Castagna09toplas},
the satisfacion of a set of constraints~\cite{Buscemi07ccpi},
or of some logical formula~\cite{Artikis09tocl,BZ10lics,PrisacariuS07formal}.
Most of the existing approaches tacitly assume that,
once a set of compliant contracts has been found, then
the services that advertised such contracts will behave
accordingly.
In other words, services are assumed to be \emph{honest},
in that they always respect the promises made.

In open
and dynamic systems, the assumption that all services are honest is
not quite realistic. In fact, services have different individual
goals, are made available by different providers, and possibly do
not trust each other.
What happens is that
services agree upon some contracts,
but may then violate them, either intentionally or not.
Since this situation may repeatedly occur in practice,
it should not be dealt with as the failure of the whole system.
Instead, contract violations should be
automatically detected and sanctioned
by the service infrastructure.

The fact that violations may be sanctioned 
gives rise to a new kind of attacks,
that exploit possible discrepancies between the 
promised and the runtime behaviour of services.
If a service does not accurately behave as promised,
an attacker can induce it to a situation
where the service is sanctioned, while the attacker
is reckoned honest.
A crucial problem is then how to avoid that a service
results \emph{culpable} of a contract violation,
despite of the honest intentions of its developer.
More formally, the problem is that of deciding
if a process \emph{realizes} a contract:
when this holds, the process is guaranteed to never
be culpable w.r.t.\ the contract
in all the possible execution contexts.  

In this paper we develop a formal theory of contract-oriented systems
that enjoys a sound criterion for establishing if a process always
realizes its contracts.  Our theory combines two basic ingredients: a
calculus of contracts, and a calculus of processes that use contracts
to interact.
Contracts are used by distributed participants to reach agreements;
once stipulated, participants can inspect them and decide what to
do next.

Ideally, a honest participant is supposed
to harmoniously evolve with her contracts;
more realistically, our theory also encompasses computations of
\emph{dishonest} participants, which may violate at run-time
some contracts they have stipulated.
A remarkable result (Theorem~\ref{lem:compliant-frown}) is that
it is always possible to detect who is culpable of a contract violation
at each state of a computation.
Also, a participant can always exculpate herself
by performing the needed actions
(Theorems~\ref{lem:compliant-smiley} and~\ref{lem:doubledo}).

Notably, instead of defining an ad-hoc model,
we have embedded the contract calculus in~\cite{Castagna09toplas}
within the process calculus \coco\!~\cite{BTZ11ice}.
To do that, the contracts of~\cite{Castagna09toplas}
have been slightly adapted to define culpability,
and \coco\ has been specialized to use these contracts.
We have formalised when a participant realizes a contract in a given
context, i.e.\ when she is never (irreparably) culpable in
computations with that context, and when she is \emph{honest}, i.e.\
when she realizes \emph{all} her contracts, in \emph{all} possible contexts.
The problem of deciding whether a participant is honest
is undecidable, in general (Theorem~\ref{th:honesty-undecidable}).
Indeed, one would have to check infinitely many contexts.
Furthermore, participants themselves are infinite state systems,
which feature recursion and parallel composition.
Our main contribution (Theorem~\ref{th:canonical-is-saint})
is a sound criterion for detecting when a participant is honest.
Technically this is achieved by defining a semantics of participants
that abstracts away the behaviour of the context.
Such semantics allows us to define when a participant
fulfills her contracts, even in the presence of dishonest participants.

\section{A calculus of contracts} \label{sect:contracts}

We assume a finite set of \emph{participant names} 
(ranged over by ${\pmv A}, {\pmv B}, \ldots$)
and a denumerable set of \emph{atoms}
(ranged over by $\atom{a}, \atom{b}, \ldots$).
We postulate an involution $\co{\atom{a}}$, also written as $\bar{\atom{a}}$,
extended to sets of atoms in the natural way.

Def.~\ref{def:contracts:syntax} introduces the syntax of contracts,
taking inspiration from~\cite{Castagna09toplas}.
We distinguish between (\emph{unilateral}) contracts $c$, which model the 
promised behaviour of a single participant, and \emph{bilateral} contracts
$\gamma$, which combine the contracts of two participants.

\begin{definition} \label{def:contracts:syntax}
\emph{Unilateral contracts} are defined by the following grammar:
\begin{align*}
    c,d \;\; & ::= \;\;
    \SumInt[i \in \mathcal{I}]{\atom{a}_i}{c_i} \ \sep \ 
    \SumExt[i \in \mathcal{I}]{\atom{a}_i}{c_i} \ \sep \
    \ready{\atom{a}}.c \ \sep \
    \rec{X}{c}
    \sep \ X
  \end{align*}
  where 
  $(i)$ the index set $\mathcal{I}$ is finite; $(ii)$ the atoms in
  $\{\atom{a}_i\}_{i \in \mathcal I}$ are pairwise distinct; $(iii)$
  the $\ready{}\!$ prefix may appear at the top-level, only; $(iv)$
  recursion is guarded.

  Let $\atom e$ be a distinguished atom such that $\atom{e}=\bar{\atom
    e}$ and whose continuation is the contract
$\E = \rec{X}{\sumI{\atom{e}}{X}}$.
  We say that \emph{$c$ succeeds} iff either $c = \sumI{\atom{e}}{\E}
  \sumInt d$, or $c = \sumE{\atom{e}}{\E} \sumExt d$, or $c =
  \ready{\atom{e}}.\; \E$.
  We will omit trailing occurrences of $\E$ in contracts.

  \emph{Bilateral contracts}
  are terms of the form $\bic c d$,
  where $\pmv A \neq \pmv B$ and at most one occurrence of
  $\ready{}\!$ is present.
\end{definition}

Intuitively, the internal sum 
$\SumInt[i \in \mathcal{I}]{\atom{a}_i}{c_i}$ 
allows to choose one of the branches $\sumI{\atom{a}_i}{c_i}$,
to perform the action $\atom{a}_i$, and then behave according to $c_i$.
Dually, the external sum  
$\SumExt[i \in \mathcal{I}]{\atom{a}_i}{c_i}$ 
constrains to wait for the other participant to choose one of the branches
$\sumE{\atom{a}_i}{c_i}$, then to perform the corresponding
$\atom{a}_i$ and finally behave according to $c_i$.
Separators $;$ and $.$ allow us to distinguish 
singleton \emph{internal} sums (\text{e.g.}, $\sumI{\atom{a}}{c}$) from
singleton \emph{external} sums (\text{e.g.}, $\sumE{\atom{a}}{c}$).
The atom $\atom{e}$ (for ``end'') enables a participant to successfully
terminate, similarly to~\cite{Castagna09toplas}. This will be reflected in
Def.~\ref{def:culpable}.
Hereafter, we shall always consider contracts with no free
occurrences of recursion variables $X$.
We shall use the binary operators to isolate a branch in a sum: 
e.g.~$(\sumI{\atom{a}}{c}) \sumInt c'$ where $c'$ is an internal sum.

The evolution of bilateral contracts 
is modelled by a labelled transition relation $\cmove{\mu}$
(Def.~\ref{def:contracts:semantics}),
where labels $\mu = {\pmv A} \says \atom{a}$ model a participant {\pmv A} 
performing the action \atom{a}.

\begin{definition} \label{def:contracts:semantics}
The relation $\cmove{\mu}$ on bilateral contracts is the smallest
relation closed 
under the rules in Fig.~\ref{fig:contracts:semantics:nonfail} 
and under the structural congruence relation $\equiv$, defined as the
least congruence which includes $\alpha$-conversion of recursion
variables, and satisfies
$\rec{X}{c} \equiv c \setenum{\bind{X}{\rec{X}{c}}}$
and
$\SumInt[i \in \mathcal{\emptyset}]{\atom{a}_i}{c_i} \equiv
    \SumExt[i \in \mathcal{\emptyset}]{\atom{a}_i}{c_i}$.
Accordingly, empty sums (either internal or external) 
will be denoted with $\cnil$.
We will \emph{not} omit trailing occurrences of $\cnil$.
Hereafter we shall consider contracts up to $\equiv$.
\end{definition}

\begin{figure}[t]
\hrulefill
\[
\begin{array}{c}
\begin{array}{rcll}
  \bic
  {(\sumI{\atom{a}}{c} \sumInt c')}
  {(\sumE{\bar{\atom{a}}}{d} \sumExt d')}
   & \cmove{{\pmv A} \says \atom{a}} &
   \bic{c}{\ready{\bar{\atom a}}.d}
   \hspace{20pt}
   & \nrule{[IntExt]}
\\[8pt]
  \bic
  {(\sumI{\atom{a}}{c} \sumInt c')}
  {\sumI{\bar{\atom{a}}}{d}}
   & \cmove{{\pmv A} \says \atom{a}} &
   \bic{c}{\ready{\bar{\atom a}}.d}
   & \nrule{[IntInt]}
\\[8pt]
  \bic
  {(\sumE{\atom{a}}{c} \sumExt c')}
  {(\sumE{\bar{\atom{a}}}{d} \sumExt d')}
   & \cmove{{\pmv A} \says \atom{a}} &
   \bic{c}{\ready{\bar{\atom a}}.d}
   & \nrule{[ExtExt]}
\\[8pt]
  \bic{\ready{a}.\ c}{d}
   & \cmove{{\pmv A} \says \atom{a}} &
   \bic{c}{d}
   & \nrule{[Rdy]}
\\[8pt]
\end{array} 
\\
\begin{array}{cl}
 \irule
  {\atom{a} \not\in \co{\setenum{\atom{b}_i}_{i \in I}}}
  {{\pmv A} \says \sumI{\atom{a}}{c} \sumInt c' \mid
   {\pmv B} \says \SumExt[i \in I]{\atom{b}_i}{d_i}
   \cmove{{\pmv A} \says \atom{a}}
   {\pmv A} \says \E \mid {\pmv B} \says \cnil}
& \nrule{[IntExtFail]}
\\[16pt]
  \irule
  {\setenum{\atom{a}} \neq \co{\setenum{\atom{b}_i}_{i \in I}}}
  {{\pmv A} \says \sumI{\atom{a}}{c} \sumInt c' \mid
   {\pmv B} \says \SumInt[i \in I]{\atom{b}_i}{d_i}
   \cmove{{\pmv A} \says \atom{a}}
   {\pmv A} \says \E \mid {\pmv B} \says \cnil}
& \nrule{[IntIntFail]}
\\[16pt]
  \irule
  {(\setenum{\atom{a}} \cup \setenum{\atom{a}_i}_{i \in I}) \;\cap\; 
   \co{\setenum{\atom{b}_i}_{i \in J}} \; = \; \emptyset}
  {{\pmv A} \says (\sumE{\atom{a}}{c} \sumExt \SumExt[i \in I]{\atom{a}_i}{c_i}) \mid
   {\pmv B} \says \SumExt[i \in J]{\atom{b}_i}{d_i}
   \cmove{{\pmv A} \says \atom{a}}
   {\pmv A} \says \E \mid {\pmv B} \says \cnil}
\;\;\;\;
& \nrule{[ExtExtFail]}
\end{array}
\end{array}
\]
\hrulefill
\vspace{-5pt}
\caption{Semantics of contracts (rules for $\pmv B$ actions omitted)}
\label{fig:contracts:semantics}
\label{fig:contracts:semantics:nonfail}
\label{fig:contracts:semantics:fail}
\vspace{-10pt}
\end{figure}

In the first three rules in Fig.~\ref{fig:contracts:semantics}, 
{\pmv A} and {\pmv B} expose complementary actions $\atom a,\bar{\atom a}$.
In rule~\nrule{[IntExt]}, participant {\pmv A} selects the branch $\atom a$ in an internal sum.
Participant {\pmv B} is then forced to commit to the corresponding branch $\bar{\atom a}$ in his external sum: this is done by marking that branch with $\ready{\bar{\atom a}}$ while discarding all the other branches. Participant {\pmv B} will then perform his action in the subsequent step, by rule~\nrule{[Rdy]}.
In rule~\nrule{[IntInt]}, both participants make an internal choice; a reaction is possible only if one of the two is a singleton --- {\pmv B} in the rule ---
namely he can only commit to his unique branch.
Were {\pmv B} exposing multiple branches, the transition would not be allowed, to account for the fact that {\pmv B} could pick a conflicting internal choice w.r.t.\ that of {\pmv A}.
In rule~\nrule{[ExtExt]}, both participants expose external sums with complementary actions, and each of the two can choose a branch (unlike in the case~\nrule{[IntExt]}, where the internal choice has to move first).
In the~\nrule{[*Fail]} rules, the action chosen by {\pmv A} is not supported by {\pmv B}.
Then, {\pmv A} will reach the success state~$\E$, while {\pmv B} will fall into the failure state~$\cnil$.

\begin{example} \label{ex:contracts:1}
Let
\(
  \gamma =
  {\pmv A} \says (\sumI{\atom{a}}{c_1} \sumInt \sumI{\atom{b}}{c_2}) \mid 
  {\pmv B} \says (\sumE{\bar{\atom{a}}}{d_1} \sumExt \sumE{\bar{\atom{c}}}{d_2})
\).
If the participant {\pmv A} internally chooses to perform \atom{a}, 
then $\gamma$ will take a transition to 
\(
  {\pmv A} \says c_1 \mid {\pmv B} \says \ready{\bar{\atom{a}}}.d_1
\).
Suppose instead that {\pmv A} chooses for perform \atom{b}, 
which is not offered by {\pmv B} in his external choice.
In this case, $\gamma$ will take a transition to 
\(
  {\pmv A} \says \E \mid {\pmv B} \says \cnil
\),
where $\cnil$ indicates that {\pmv B} 
cannot proceed with the interaction.
Coherently with~\cite{Castagna09toplas}, 
below we will characterise this behaviour by saying that
the contracts of {\pmv A} and {\pmv B} are \emph{not} compliant.
\end{example}


The following lemma states that bilateral contracts are never stuck
unless both participants have contract $\cnil$.  
Actually, if none of the first four rules in Fig.~\ref{fig:contracts:semantics}
can be applied, the contract can make a transition 
with one of the~\nrule{[*Fail]} rules.

\newcommand{\lemstuckiffnil}[0]
{
A bilateral contract $\bic{c}{d}$ is stuck iff $c = d = \cnil$.
}
\begin{lemma} \label{lem:stuck-iff-nil}
\lemstuckiffnil
\end{lemma}


Below we establish that contracts are deterministic.  This is
guaranteed by the requirement $(ii)$ of
Def.~\ref{def:contracts:syntax}.  
Determinism is a very desirable property indeed,
because it ensures that the duties of a participant at any
given time are uniquely determined by the past actions.
Note that the contracts in~\cite{Castagna09toplas} satisfy
distributivity laws like $\sumI{\atom a}{c} \sumInt \sumI{\atom a}{d}
= \sumI{\atom a}{c \sumInt d}$, which allow for rewriting them so that
$(ii)$ in Def.~\ref{def:contracts:syntax} holds.  Therefore, $(ii)$ is
not a real restriction w.r.t.~\cite{Castagna09toplas}.

\newcommand{\lemcontractsdeterminism}
{
For all $\gamma$, if $\gamma \cmove{\mu} \gamma'$ and $\gamma \cmove{\mu} \gamma''$,
then $\gamma' = \gamma''$. 
}
\begin{lemma}[Determinism] \label{lem:contracts:determinism}
\lemcontractsdeterminism
\end{lemma}

\vspace{-10pt}
\paragraph{Compliance.}

Below we define when two contracts are \emph{compliant}, in a similar
fashion to~\cite{Castagna09toplas}.  Intuitively, two contracts are
compliant if whatever sets of choices they offer, there is at least
one common option that can make the contracts progress.
Differently from~\cite{Castagna09toplas}, our notion of compliance is
symmetric, in that we do not discriminate between the participant
roles as client and server.  
Consequently, we do not consider compliant
two contracts where only one of the parties is willing to terminate.
For example, the buyer contract $\sumI{\atom{ship}}{\E}$ is not
compliant with the seller contract
$\sumE{\overline{\atom{ship}}}{\sumI{\atom{pay}}{\E}}$, because the
buyer should not be allowed to terminate if the seller still requires
to be paid.

Similarly to~\cite{Castagna09toplas}, given two contracts we observe
their \emph{ready sets} (Def.~\ref{def:rs}) to detect when the
enabled actions allow them to synchronise correctly.

\begin{definition}[Compliance] \label{def:compliance} \label{def:rs}
For all contracts $c$, we define the set of sets $\rs{c}$ as:
\vspace{-2pt}
\[
\begin{array}{c}
  \rs{\cnil} = \setenum{\emptyset}
  \hspace{20pt}
  \rs{\ready{\atom{a}}.c} = \setenum{\setenum{\ready{\!}}}
  \hspace{20pt}
  \rs{\rec{X}{c}} = \rs{c} 
\\[5pt]
  \rs{\SumInt[i \in I]{\atom{a}_i}{c_i}} = 
  \setcomp{\setenum{\atom{a}_i}}{i \in I} 
  \textit{ if } I \neq \emptyset
  \hspace{20pt}
  \rs{\SumExt[i \in I]{\atom{a}_i}{c_i}} = 
  \setenum{\setcomp{\atom{a}_i}{i \in I}} 
  \textit{ if } I \neq \emptyset
\end{array}
\]
The relation $\compliant$ between contracts is the largest relation
such that, whenever $c \compliant d$:
\begin{itemize}

\item[(1)] \( 
\forall \mathcal{X} \in \rs{c}, \mathcal{Y} \in \rs{d}.\ \co{\mathcal{X}} \cap \mathcal{Y} \neq \emptyset 
\;\text{ or }\; \ready{} \in (\mathcal{X} \cup \mathcal{Y}) \setminus (\mathcal{X} \cap \mathcal{Y})
\)

\item[(2)] \(
{\pmv A} \says c \mid {\pmv B} \says d \cmove{\mu}
{\pmv A} \says c' \mid {\pmv B} \says d' \implies c' \compliant d'
\)

\end{itemize}
When $c \compliant d$, we say that the contracts $c$ and $d$ are 
\emph{compliant}.
\end{definition}

\begin{example} \label{ex:contracts:2}
Recall from Ex.~\ref{ex:contracts:1} the contracts 
$c = \sumI{\atom{a}}{c_1} \sumInt \sumI{\atom{b}}{c_2}$ and
$d = \sumE{\bar{\atom{a}}}{d_1} \sumExt \sumE{\bar{\atom{c}}}{d_2}$.
We have that $\rs{c} = \setenum{\setenum{\atom{a}},\setenum{\atom{b}}}$, and
$\rs{d} = \setenum{\setenum{\bar{\atom{a}}, \bar{\atom{c}}}}$,
which do not respect item $(1)$ of Def.~\ref{def:compliance}
(take $\mathcal{X} = \setenum{\atom{b}}$ and
$\mathcal{Y} = \setenum{\bar{\atom{a}}, \bar{\atom{c}}}$).
Therefore, $c$ and $d$ are \emph{not} compliant.
\end{example}


The following lemma provides an alternative characterization of compliance.
Two contracts are compliant iff, when combined into a bilateral contract
$\gamma$, no computation of $\gamma$ reaches a state where one
of the contracts is  $\cnil$.
Together with Lemma~\ref{lem:stuck-iff-nil}, we have that such $\gamma$
will never get stuck.

\newcommand{\lemcompliantfail}[0]{
For all bilateral contracts $\gamma = {\pmv A} \says c \mid {\pmv B} \says d$: 
\[
  c \compliant d
  \iff
  \big(
  \forall c',d' .\ 
  \gamma
  \cmove{}^* 
  {\pmv A} \says c' \mid {\pmv B} \says d'
  \implies
  c' \neq \cnil \text{ and } d' \neq \cnil
  \big)
\]
}
\begin{lemma} \label{lem:compliant-fail}
\lemcompliantfail
\end{lemma}


The following lemma guarantees, for all $c$ not containing $\cnil$, 
the existence of a contract $d$ compliant with $c$.
Intuitively, we can construct $d$ from $c$ by 
turning internal choices into external ones (and \emph{viceversa}),
and by turning actions into co-actions.

\newcommand{\lemcompliantexists}[0]{
For all $\cnil$-free contracts $c$, there exists $d$ 
such that $c \compliant d$.
}
\begin{lemma} \label{lem:compliant-exists}
\lemcompliantexists
\end{lemma}

\paragraph{Culpability.}
We now tackle the problem of determining who is expected
to make the next step for the fulfilment of a bilateral contract.
We call a participant {\pmv A} \emph{culpable} in $\gamma$ if
she is expected to perform some action so to make $\gamma$ progress.
Also, we consider {\pmv A} culpable when she is advertising 
the ``failure'' contract $\cnil$. 
This agrees with our \nrule{[*Fail]} rules, which set
{\pmv A}'s contract to $\cnil$ when the other participant legitimately chooses
an action not supported by {\pmv A}. 
Note that we do not consider {\pmv A} culpable when her contract has
enabled $\atom{e}$ actions. 

\begin{definition} \label{def:culpable}
A participant $\pmv A$ is culpable in $\gamma = \bic{c}{d}$, written
$\cfrown{\pmv A}{\gamma}$, iff:
\[
  c = \cnil
  \quad\lor\quad
  \big(
  \gamma \not\cmove{{\pmv A} \says \atom{e}}
  \;\;\land\;\;
  \exists \atom{a}.\;\;
  \gamma \cmove{{\pmv A} \says \atom{a}}
  \big)
\]
When $\pmv A$ is \emph{not} culpable in $\gamma$ 
we write $\csmiley{\pmv A}{\gamma}$.
\end{definition}

The following result states that a participant {\pmv A} is always able
to recover from culpability by performing some of her
duties. Furthermore, this requires at most two steps in an 
``{\pmv A}-solo'' trace where no other participant intervenes.

\begin{definition}\label{def:solo}
Let $\xrightarrow{}$ be an LTS with
labels of the form ${\pmv A}_i \says (\cdots)$, 
for ${\pmv A}_i$ ranging over participants names.
For all {\pmv A}, 
we say that a $\xrightarrow{}$-trace $\eta$
is $\pmv A$-solo iff 
$\eta$ only contains labels of the form ${\pmv A} \says (\cdots)$.
If $\eta = (\mu_i)_{i \in 0..n}$, we will write
$\xrightarrow{\eta}$ for \mbox{$\xrightarrow{\mu_0}\cdots\xrightarrow{\mu_n}$}.
\end{definition}

\newcommand{\lemcompliantsmiley}{
For all $\gamma = \bic{c}{d}$ with $\cnil$-free $c$,
there exists $\gamma'$ and {\pmv A}-solo $\eta$ with $|\eta| \leq 2$
such that $\gamma \cmove{\eta} \gamma'$
and $\csmiley{\pmv A}\gamma'$.
}
\begin{theorem}[Contractual exculpation]\label{lem:compliant-smiley}
\lemcompliantsmiley
\end{theorem}

A crucial property of culpability is to ensure that
either two participants are both succeeding, 
or it is possible to single out who has to make the next step.
An external judge is
therefore always able to detect who is violating the contracts agreed upon.

\newcommand{\lemcompliantfrown}{
For all $c, d$ 
if $c \compliant d$ and 
$\bic{c}{d}\cmove{}^* \gamma = \bic{c'}{d'}$, 
then either 
$c'$ and $d'$ succeed, 
or $\cfrown{\pmv A}{\gamma}$,
or $\cfrown{\pmv B}{\gamma}$.
}
\begin{theorem} \label{lem:compliant-frown}
\lemcompliantfrown
\end{theorem}

\begin{example} \label{ex:success-culpable}
A participant might be culpable even though her contract succeeds.
For instance, let $\gamma = \bic{c}{d}$, where
$c = \atom{e} \sumExt \bar{\atom{a}}$ and $d = \atom{a} \sumExt \atom{b}$.
By Def.~\ref{def:contracts:syntax} we have that $c$ succeeds,
but {\pmv A} is culpable in $\gamma$ because
she cannot fire $\atom{e}$, while she can fire $\bar{\atom{a}}$
by rule~\nrule{[ExtExt]}.
This makes quite sense, because {\pmv A} is saying that
she is either willing to terminate or to perform $\bar{\atom{a}}$,
but the other participant is not allowing {\pmv A} to terminate.
Note that also {\pmv B} is culpable, because he can fire $\atom{a}$.
\end{example}

\section{A Calculus of Contracting Processes}\label{sect:co2}

We now embed the contracts introduced in \S~\ref{sect:contracts}
in a specialization of the parametric process calculus
\coco~\cite{BTZ11ice}.
Let $\vars$ and $\snames$ be two disjoint countably infinite sets
of \emph{session variables} (ranged over by $x,y,\ldots$) and 
\emph{session names} (ranged over by $s,t,\ldots$).
Let $u,v,\ldots$ range over $\vars \cup \snames$.


\begin{definition}\label{def:co2:syntax}
The abstract syntax of \coco is given by the following productions:
\[
\begin{array}{rcccccccccccc}
    \text{Systems}   \qquad S  & ::= & 
    \emptysys &\sep& \sys {\pmv A} P &\sep& \sys s \gamma &\sep& S \mid S &\sep& (u)S
    \\[.3pc] 
    \text{Processes} \qquad P  & ::=  & \freeze u {\pmv A \says c}
    &\sep&    \textstyle \sum_{i} \pi_i.P_i
    &\sep&    P \mid P
    &\sep&    (u)P 
    &\sep&    X(\vec u)
    \\[.3pc] 
    \text{Prefixes}  \qquad \pi
    & ::= & \tau
    &\sep&    \tell {\pmv A} {\freeze u c}
    &\sep&    \fuse u 
    &\sep&    \fact u {\atom{a}}
    &\sep&    \ask {u} \phi
\end{array}
\]
\end{definition}

The only binder for session variables and names is the
delimitation (both in systems and processes). 
Free variables/names are defined accordingly, 
and they are denoted by $\fv\_$ and $\fn\_$.
A system or a process is \emph{closed} when it has no free variables.

Systems are the parallel composition of 
\emph{participants} $\sys {\pmv A} P$
and \emph{sessions} $\sys s \gamma$.

\begin{figure}[t]
  \hrulefill
  \small
\begin{center}
commutative monoidal laws for $\mid$ on processes and systems
\end{center}
\vspace{-10pt}
\[
\begin{array}{c}
  \sys u {(v)P} \equiv \sys{(v) \, u} P 
  \;\;\text{if}\ u \neq v
  \hspace{10pt}
  Z \mid (u)Z' \equiv (u)(Z \mid Z') 
  \;\;\text{if}\ u \not\in \fv Z \cup \fn Z
  \hspace{10pt}
  (u)(v)Z \equiv (v)(u)Z
  \\[8pt]
  (u)Z \equiv Z
  \;\;\text{if}\ u \not\in \fv Z \cup \fn Z
  \hspace{20pt}
  \sys{\pmv A}{K} \mid \sys{\pmv A}{P} 
  \equiv 
  \sys {\pmv A} {K \mid P}
  \hspace{20pt} 
  \freeze s c \equiv \pnil \equiv \fuse s .\, P
\end{array}
\]
\hrulefill
\vspace{-5pt}
\caption{Structural equivalence for \coco ($Z,Z'$ range over systems or
  processes)} \label{fig:co2:equiv}
\vspace{-10pt}
\end{figure}

A \emph{latent contract} $\freeze x {\pmv A \says c}$ 
represents a contract $c$ (advertised by {\pmv A}) which
has not been stipulated yet; upon stipulation, $x$ will be
instantiated to a fresh session name.
We impose that in a system $\sys{\pmv A}{P} \mid \sys{\pmv A}{Q} \mid
S$, either $P$ or $Q$ is a parallel composition of latent contracts.
Hereafter, $K,K',\ldots$
are meta-variables for compositions of latent contracts.
We allow prefix-guarded finite sums of processes,
and write $\pi_1.P_1 + \pi_2.P_2$ for $\sum_{i=1,2}\pi_i.P_i$,
and $\pnil$ for $\sum_{\emptyset}P$. 
Recursion is allowed only for processes; for this we stipulate that
each process identifier $X$ has a unique defining equation
$X(u_1, \ldots, u_j) \mmdef P$ such that $\fv{P} \subseteq
\{u_1,\ldots,u_j\} \subseteq \vars$ and each occurrence of process
identifiers in $P$ is prefix-guarded.

Prefixes include silent action $\tau$, 
contract advertisement $\tell{\pmv A}{\freeze u c}$,
contract stipulation $\fuse u$,
action execution $\fact{u}{\atom a}$, and 
contract query $\ask{u}{\phi}$.
In each prefix $\pi \neq \tau$, $u$ refers to the target session involved in
the execution of $\pi$.
We omit trailing occurrences of $\pnil$.

Note that participants can only contain latent contracts,
while sessions can only contain bilateral contracts, constructed from
latent contracts upon reaching agreements.

  
The semantics of \coco\ is formalised by a reduction relation $\pmove{}$
on systems that relies on the structural congruence defined in
Fig.~\ref{fig:co2:equiv}, where the last law allows for collecting
garbage terms possibly arising from variable substitutions.
\begin{definition}\label{def:co2:semantics}
  The relation $\pmove{}$ is the smallest relation closed under the
  rules of Fig.~\ref{fig:co2:semantics}, defined over systems up to
  structural equivalence, as defined in Fig.~\ref{fig:co2:equiv}.
  The relation $\agreement K {(\gamma, K')} x \sigma$ holds iff
  $(i)$ $K$ has the form $\freeze y {\pmv A \says c} \mid \freeze z
  {\pmv B \says d} \mid K'$, $(ii)$ $c \compliant d$, $(iii)$ $\gamma
  = \bic{c}{d}$, and $(iv)$ $\sigma = \setenum{\bind{x,y,z} s}$ maps all $x,y,z \in \vars$ to $s \in \names$.
\end{definition}
\begin{figure}[t]
\hrulefill
  \[
  \begin{array}{cl}
    {\sys {\pmv A} {\tau . P + P' \mid Q}
      \pmove{}
      \sys {\pmv A} {P \mid Q}
    } & \nrule{[Tau]}
    \\[10pt]
    {\sys {\pmv A} {\tell {\pmv B} {\freeze x c} . P + P' \mid Q}
      \pmove{}
      \sys {\pmv A} {P \mid Q} \ \mid\ 
      \sys {\pmv B} {\freeze x {\pmv A \says c}}
    } & \nrule{[Tell]}
    \\[10pt]
    \irule
    {\agreement K {(\gamma,K')} x \sigma
      \qquad \vec u = \dom \sigma
      \qquad s = \sigma(x) \;\; \text{fresh}
    }
    {(\vec u)(\sys {\pmv A} {\fuse x . P + P'  \mid K \mid Q} \mid S)
      \pmove{}
      (s)(\sys {\pmv A} {P \mid Q \mid K'}\sigma \ \mid\ \sys s {\gamma} \mid S\sigma)
    } \hspace{15pt} & \nrule{[Fuse]}
    \\[14pt]
    \irule
    {\gamma \cmove{\pmv A \says \atom{a}} \gamma'}
    {\sys s {\gamma} \ \mid \ 
      \sys {\pmv A} {\fact s {\atom{a}} . P + P' \mid Q}
      \pmove{} 
      \sys s {\gamma'} \ \mid \ 
      \sys {\pmv A} {P \mid Q}
    } & \nrule{[Do]}
    \\[15pt]
    \irule
    {\gamma \vdash \phi}
    {\sys {\pmv A} {\ask {s} \phi . P + P'  \mid Q} \mid \sys s \gamma
      \pmove{}
      \sys {\pmv A} {P \mid Q} \mid \sys s \gamma
    } & \nrule{[Ask]}
\end{array}
\]
\vspace{0pt}
\[
\begin{array}{c}
  \irule
    {X(\vec u) \mmdef P \qquad P\subs{\vec v}{\vec u} \pmove{} P'}
    {X(\vec v) \pmove{} P'}
    \; \nrule{[Def]}
    \hspace{20pt}
    \irule
    {S \pmove{} S'}
    {S \mid S'' \pmove{} S' \mid S''}
    \; \nrule{[Par]}
    \hspace{20pt}
    \irule
    {S \pmove{} S'}
    {(u)S \pmove{} (u)S'}
    \; \nrule{[Del]}
  \end{array}
\]
\hrulefill
\vspace{-5pt}
\caption{Reduction semantics of \coco} \label{fig:co2:semantics}
\vspace{-10pt}
\end{figure}

Rule~\nrule{[Tau]} simply fires a $\tau$ prefix as expected.
Rule~\nrule{[Tell]} advertises a latent contract 
$\freeze x {\pmv A \says c}$, 
by putting it in parallel with the existing 
participants and sessions (the structural congruence laws in Fig.~\ref{fig:co2:equiv} allow for latent contracts to float in a system and, by the second last law, to move across the boxes of participants as appropriate).
Rule \nrule{[Fuse]} finds agreements among the latent contracts $K$ of
$\pmv A$; an agreement is reached when $K$ contains a bilateral
contract $\gamma$ whose unilater contracs are compliant
(cf. Def.~\ref{def:co2:semantics}). 
Note that, once the agreement is reached, 
the compliant contracts start a fresh session containing $\gamma$.
Rule \nrule{[Do]} allows a participant $\pmv A$ to fulfill her
contract $\gamma$, by performing the needed actions in the session
containing $\gamma$ (which, accordingly, evolves to $\gamma'$).
Rule \nrule{[Ask]} checks if a condition $\phi$ holds in a session.
The actual nature of $\phi$ is almost immaterial in this paper: the reader
may assume that $\phi$ is a formula in an LTL
logic~\cite{Emerson90temporal}. 
For closed $\gamma$ and $\phi$, 
$\gamma \vdash \phi$ holds iff $\gamma \models_{LTL} \phi$ according to the
standard LTL semantics where, for a $\cmove{}$-trace $\eta = (\gamma_i
\cmove{\mu_i} \gamma_{i+1})_i$ from $\gamma_0 = \gamma$, we define
\(
  \eta \models \atom a
  \iff
  \exists \pmv A .\; \mu_0 = \pmv A \says {\atom a}
\).
The last three rules are standard.

Hereafter it will be sometimes useful to record 
the prefix $\pi$ fired by {\pmv A}
by implicitly decorating the corresponding reduction step, 
as in $\pmove{{\pmv A} \says \pi}$.

\medskip

The rest of this section is devoted to a few examples that highlight how
bilateral contracts can be used in \coco\!.

\begin{example} \label{ex:store:co2}
Consider an online store {\pmv A} with the following contract $c_{\pmv A}$:
buyers can add items to the shopping cart, 
and then either leave the store or pay with a credit card.
Assume the store modelled as the \coco process
$P_{\pmv A} = (x) \, (\tell{\pmv A}{\freeze{x}{c_{\pmv A}}}.\, X \mid \fuse{x}{})$, 
where:
\begin{align*}
  c_{\pmv A} & = \rec Z {
  \sumE{\atom{addToCart}}{Z} \sumExt 
  \sumE{\atom{creditCard}}
       {(\overline{\atom{ok}} \sumInt \overline{\atom{no}})}
  \sumExt \atom{e}
  } \\
  X & \mmdef 
  \fact{x}{\atom{addToCart}}.X + 
  \fact{x}{\atom{creditCard}}. 
  (\tau. \fact{x}{\overline{\atom{ok}}} + \tau.\fact{x}{\overline{\atom{no}}}) 
\end{align*}
Let {\pmv B} be a buyer with contract 
\(
  c_{\pmv B} = \sumI{\overline{\atom{addToCart}}}{
    \sumI{\overline{\atom{creditCard}}}
    {
      {(\atom{ok} \sumExt \atom{no})}
    }
  }
\),
and let:
\[
  P_{\pmv B} =
  (y) \, \tell{\pmv A}{\freeze{y}{c_{\pmv B}}}.\, Y
  \hspace{50pt}
  Y \mmdef 
  \fact{y}{\overline{\atom{addToCart}}}.\, 
  \fact{y}{\overline{\atom{creditCard}}}.\,
  \fact{y}{\atom{ok}}
\]
A possible, successful, computation of the system 
$S = \sys{\pmv A}{P_{\pmv A}} \mid \sys{\pmv B}{P_{\pmv B}}$ 
is the following:
\begin{align*}
\small
  S \pmove{}^* 
  & (x,y) \; \big(
  \sys {\pmv A} {\freeze{x}{{\pmv A} \says c_{\pmv A}} \mid
    \freeze{y}{{\pmv B} \says c_{\pmv B}} \mid \fuse{x}{} \mid X}
  \mid
  \sys {\pmv B} {Y} 
  \big) \\
  \pmove{}\hspace{4pt}
  & (s)\; \big(
  \sys {\pmv A} {X\setenum{\bind{x}{s}}}
  \mid
  \sys {\pmv B} {Y\setenum{\bind{y}{s}}}
  \mid 
  \sys s {\bic{c_{\pmv A}}{c_{\pmv B}}} 
  \big) \\
  \pmove{}^* 
  & (s)\; \big(
  \sys {\pmv A} {X\setenum{\bind{x}{s}}}
  \mid
  \sys {\pmv B} {\fact{s}{\overline{\atom{creditCard}}}.\,
  \fact{y}{\atom{ok}}
  }
  \mid
  \sys s {\bic{c_{\pmv A}}{\sumI{\overline{\atom{creditCard}}}
    {
      {(\atom{ok} \sumExt \atom{no})}
    }}}
  \big) \\
  \pmove{}^* 
  & (s)\; \big(
  \sys {\pmv A} {\tau. \fact{x}{\overline{\atom{ok}}} + \tau.\fact{x}{\overline{\atom{no}}}}
  \mid
  \sys {\pmv B} {\fact{y}{\atom{ok}}}
  \mid 
  \sys s {\bic
    {\overline{\atom{ok}} \sumInt \overline{\atom{no}}}
    {\atom{ok} \sumExt \atom{no}}}
  \big) \\
  \pmove{} \hspace{4pt}
  & (s)\; \big(
  \sys {\pmv A} {\fact{x}{\overline{\atom{ok}}}}
  \mid
  \sys {\pmv B} {\fact{y}{\atom{ok}}}
  \mid 
  \sys s {\bic
    {\overline{\atom{ok}} \sumInt \overline{\atom{no}}}
    {\atom{ok} \sumExt \atom{no}}}
  \big) \\
  \pmove{}^* 
  &  (s)\; \big(  
  \sys {\pmv A} {\pnil}
  \mid
  \sys {\pmv B} {\pnil}
  \mid 
  \sys s {\bic{\E}{\E}}
  \big)
\end{align*}
\end{example}

\begin{example}\label{ex:voucher:co2}
  \newcommand{\cp}{\atom{clickPay}}
  \newcommand{\cv}{\atom{clickVoucher}}
  \newcommand{\p}{\atom{pay}}
  \newcommand{\vou}{\atom{voucher}}
  \newcommand{\rj}{\overline{\atom{reject}}}
  \newcommand{\ac}{\overline{\atom{accept}}}
  \newcommand{\ctc}{\atom{clickPay}.\p \sumExt \cv.(\rj;\p \sumInt \ac;\vou)}
  \newcommand{\ctv}{\atom{ok} \sumExt \atom{no}}
  An on-line store \pmv A offers buyers two options: $\cp$
  or $\cv$.
  If a buyer \pmv B chooses $\cp$, \pmv A accepts the payment (\p)
  otherwise \pmv A checks the validity of the voucher with \pmv V, an
  electronic voucher distribution and management system.
  If \pmv V validates the voucher, \pmv B can use it ($\vou$),
  otherwise he will pay.

  The contracts $c_{\pmv A} = \ctc$ and $c'_{\pmv A} = \ctv$ model the
  scenario above.
  A \coco\ process for \pmv A can be the following
  \begin{eqnarray*}
  P_{\pmv A} & = & (x)(\tell{\pmv A}{\freeze x {c_{\pmv A}}}.(
       \fact x {\atom{clickPay}} . \fact x \p
     + \fact x \cv . ((y) \tell{\pmv V} {\freeze y {c'_{\pmv A}}}.X)))
  \\
  X & = & \fact y {\atom{ok}}. \fact x \ac . \fact x \vou
  +   \fact y {\atom{no}}. \fact x \rj . \fact x \p
  +   \tau. \fact x \rj. \fact x \p
  \end{eqnarray*}

  Contract $c_{\pmv A}$ (resp. $c'_{\pmv A}$) is stipulated when $(i)$
  \pmv B (resp. \pmv V) advertises to \pmv A (resp. \pmv V) a contract
  $d$ with $c_{\pmv A} \compliant d$ (resp.  $c'_{\pmv A} \compliant
  d$) and $(ii)$ a $\fuse{z}{}$ is executed in \pmv A (resp. \pmv V).

  Variables $x$ and $y$ in $P_{\pmv A}$ correspond to two separate
  sessions, where \pmv A respectively interacts with \pmv B and \pmv
  V.
  The semantics of \coco\ ensures that $x$ and $y$ will be instantiated
  to different session names (if at all).

  The advertisement of $c'_{\pmv A}$ causally depends on the
  stipulation of the contracts of \pmv A and \pmv B, otherwise \pmv A
  cannot fire $\fact x \cv$.
  Instead, \pmv A and \pmv B can interact regardless the presence of
  \pmv V since $\tell{\pmv V} {\freeze y {c'_{\pmv A}}}$ is non
  blocking and the $\tau$-branch of \pmv A in $X$ is enabled (letting
  \pmv A to autonomously reject the voucher, e.g. because \pmv B is not
  entitled to use it).
\end{example}

\begin{example} \label{ex:travel:co2}
Consider a travel agency {\pmv A} which queries in parallel
an airline ticket broker {\pmv F} and a hotel reservation service {\pmv H}
in order to complete the organization of a trip.
The travel agency service $\sys {\pmv A} P$ can be defined as follows:
\newcommand{\cairline}{\freeze{x}{\sumI{\atom{ticket}}{(\atom{commitF} \sumInt \atom{abortF})}}}
\newcommand{\chotel}{\freeze{y}{\sumI{\atom{hotel}}{(\atom{commitH} \sumInt \atom{abortH})}}}
\begin{align*}
  P & = (x,y) (\tell{\pmv F}{\cairline}.X \, \mid \, \tell{\pmv H}{\chotel}.Y) \\
  X & \mmdef \fact{x}{\atom{ticket}}.\, ((\ask{y}{\it true}.\, \fact{x}{\atom{commitF}}) + \tau.\fact{x}{\atom{abortF}}) \\
  Y & \mmdef \fact{y}{\atom{hotel}}.\, ((\ask{x}{\it true}.\, \fact{y}{\atom{commitH}}) + \tau.\fact{y}{\atom{abortH}})
\end{align*}
where the $\tau$ actions model timeouts used to ensure progress.  The
travel agency in process $X$ starts buying a ticket, and commits to it
only when the hotel reservation session $y$ is started. Similarly
for process $Y$.
\end{example}

The next example shows a peculiar use of $\ask{}{}$ whereby a
participant inspects a stipulated contract
to decide its future behaviour. 

\begin{example}\label{ex:choice-via-ask}
  \newcommand{\abort}{\overline{\atom{abort}}}  
  \newcommand{\commit}{\overline{\atom{commit}}}
  \newcommand{\ccard}{\atom{creditCard}}  
  \newcommand{\bank}{\atom{bankTransfer}}
  An online store \pmv A can choose whether to abort a transaction
  ($\abort$) or to commit to the payment ($\commit$).
  In the latter case, the buyer has two options, either he
  pays by credit card ($\ccard$) or by bank transfer ($\bank$).
  The contract of {\pmv A} is modelled as 
  $c = \abort \sumInt \commit;(\ccard \sumExt \bank)$.
  Consider the process
  \[
  P_{\pmv A} = (x)(
  \tell{\pmv A}{\freeze x c}.\,
  (\ask x \phi.\,
  \fact x \commit.\, 
  \fact x \ccard
  +
  \fact x {\abort} ) )
  \]
  where $\phi = \Box (\commit \imp \neg\Diamond \bank)$.
  The process $P_{\pmv A}$ first advertises $c$.
  Once a session $\sys{s}{\gamma}$ is initiated with $\gamma = \bic{c}{d}$, 
  $\pmv A$ tests $\gamma$ through $\ask{x}{\phi}$ before
  committing to the payment. 
  If $\ask{x}{\phi}$ detects that 
  {\pmv B} has promised
  not to use the bank transfer option, then {\pmv A} commits to the
  payment, and then never offers {\pmv B} to perform a bank transfer. 
  Otherwise, if $d$ does not rule out the bank transfer,
  even if {\pmv B} might actually pay by credit card, 
  $\pmv A$ aborts the session. 
  Note that in both cases {\pmv A} realizes her own
  contract, even if she is never performing the bank transfer. 
\end{example}

\section{On honesty}\label{sect:honesty}

In this section we set out when a participant {\pmv A} 
is honest (Def.~\ref{def:honest}).
Intuitively, we consider all the possible runs of all possible systems, 
and require that in every session {\pmv A} 
is not definitely culpable.
To this aim, we first provide \coco with the counterpart
of the (non)culpability relation introduced in Def.~\ref{def:culpable}.
Intuitively, we write $\csmiley[s]{\pmv A}{S}$ when, in the system $S$, 
if the participant {\pmv A} is involved in the session $s$, 
then she is not culpable w.r.t.\ the contract stipulated therein.

\begin{definition} \label{def:co2-culpable}
We write $\csmiley[s]{\pmv A}{S}$ whenever
\(
\forall \vec{u},\gamma,S'.\; 
\big(
S \equiv (\vec{u})(\sys{s}{\gamma} \mid S')
\implies \csmiley{\pmv A}{\gamma}
\big)
\).
We write $\csmiley{\pmv A}{S}$ whenever $\csmiley[s]{\pmv A}{S}$
for all session names $s$.
\end{definition}

A technical issue is that a participant could not get a chance to
act in all the traces.
For instance, let 
$S = \sys {\pmv A}{\fact s {\atom{pay}}} \mid \sys {\pmv B}{X} \mid S'$, 
where $S'$ enables 
{\pmv A}'s action and $X \defeq \tau.X$; note that $S$ generates the infinite
trace $S \pmove{} S \pmove{} S \pmove{} \cdots$ in
which \pmv A never pays, despite her honest intention.
To account for this fact, we will check the honesty of a participant in
{\em fair} traces, only, i.e.~those where
persistent transitions are eventually followed.

\begin{definition}\label{def:fairness}\label{def:taufairness}
Given an LTS $\xrightarrow{\mu}$, we say that a (finite or infinite) 
trace $\eta = (P_i \xrightarrow{\mu_i} P_{i+1})_i$ 
having length $|\eta| \in \mathbb{N}\cup\setenum{\infty}$ 
is {\em fair}
w.r.t.~a set of labels $\mathcal{L}$ if and only if 
\[
\forall i \in \mathbb{N},\mu \in \mathcal{L}.  
\Big(
 i \leq |\eta| \land
 (\forall j \in \mathbb{N}.\;
   i \leq j \leq |\eta| \implies  P_j \xrightarrow{\mu}) \implies
 \exists j\geq i .\; \mu_j = \mu
\Big)
\]
A fair trace is a trace which is fair w.r.t.~all the labels in the LTS.
\end{definition}
Note that, by Def.~\ref{def:fairness}, a fair trace is also a maximal one
(w.r.t.~$\mathcal{L}$). Indeed, if a fair trace is finite, the condition
above guarantees that its final state has no $\mathcal{L}$ transitions
enabled. 

Finally, when checking the fairness of a trace, we shall implicitly
assume that the labels $\mu$ in our LTSs of contracts and processes
always distinguish between different occurrences of the same prefix.
E.g., a $\pmove{}$-fair trace of $\sys{\pmv A}{X \mid X}$ where $X
\mmdef \tau . X$ is not allowed to only perform the $\tau$'s of the first
$X$. Technically, labels $\mu$ always implicitly carry the syntactic
{\em address} of the prefix which is being fired, in the spirit of the
Enhanced Structured Operational Semantics~\cite{Degano01eos}.

In a \emph{stable} trace the identity of names and variables cannot be
confused by \mbox{$\alpha$-conversion}.  Indeed,
\mbox{$\alpha$-conversion} is only needed to make delimitations fresh
when unfolding recursive processes.  
W.l.o.g.\ hereafter we shall often consider stable traces, only: 
in this way we ensure that
e.g.~a name $s$ represents the same session throughout the
whole trace.

\begin{definition}\label{def:stable}
A \emph{stable} $\pmove{}$-trace is a trace
\(
(\vec{u}_0) S_0 \pmove{} (\vec{u}_1) S_1 \pmove{} 
(\vec{u}_2) S_2 \pmove{} \cdots
\)
in which 
$(1)$ all delimitations carry distinct names and variables,
$(2)$ delimitations have been brought to the top-level as
much as possible (using $\equiv$), and 
$(3)$ no $\alpha$-conversion is
performed in the trace except when unfolding recursive processes.
\end{definition}

Below, we define several notions of contract faithfulness for
participants.  
We start by clarifying when a participant $\pmv A$ {\em realizes} 
a contract (inside a session $s$) within a specific context. 
This happens when from any reachable system state $S_0$,
participant $\pmv A$ will eventually perform actions to
exculpate herself (in $s$). 
In this phase, {\pmv A} is protected from interference with other participants.  
Then, we say {\pmv A} {\em honest in a system} if she realizes every
contract in that system. 
When $\sys{\pmv A}{P}$ is honest
independently of the system, we simply say that
$\sys{\pmv A}{P}$ is {\em honest}. 
In this last case, we rule out
those systems carrying stipulated or latent contracts of
$\pmv A$ outside of $\sys{\pmv A}{P}$; 
otherwise the system can trivially make $\pmv A$ culpable:
\text{e.g.}, we disallow
$\sys{\pmv A}{P} \mid \sys{\pmv B}{ \freeze x {\pmv A \says
 \overline{\atom{pay}}} \mid \cdots }$.

\begin{definition}[Honesty] \label{def:honest} \label{def:saint}
We say that:
\begin{itemize}

\item $\pmv A$ \emph{realizes $c$ at $s$ in $S$} iff
whenever $S = (\vec{u}) ( s[\bic{c}{d}] \mid S' )$,
$S \pmove{}^* S_0$, and $(S_i)_i$ is a
$\setenum{\pmv A \says \pi }$-fair {\pmv A}-solo 
stable $\pmove{}$-trace then 
$\csmiley[s]{\pmv A}{S_j}$ for some $j \geq 0$;

\item $\pmv A$ is \emph{honest in $S$} iff for all $c$ and $s$,
${\pmv A}$ realizes $c$ at $s$ in $S$;

\item $\sys{\pmv A}{P}$ is \emph{honest} iff
for all $S$ with no $\pmv A \says \cdots$ nor $\sys{\pmv A}{\cdots}$, 
${\pmv A}$ is honest in $\sys{\pmv A}{P} \mid S$.

\end{itemize}
\end{definition}

\begin{example} \label{ex:honest:1}
A computation of the store-buyer system 
$S = \pmv A[P_{\pmv A}] \mid \pmv B[P_{\pmv B}]$ 
from Ex.~\ref{ex:store:co2} is:
\begin{align*}
\small
  S \pmove{}^* 
  & (s)\; \big(
  \sys {\pmv A} {\tau. \fact{x}{\overline{\atom{ok}}} + \tau.\fact{x}{\overline{\atom{no}}}}
  \mid
  \sys {\pmv B} {\fact{y}{\atom{ok}}}
  \mid 
  \sys s {\bic
    {\overline{\atom{ok}} \sumInt \overline{\atom{no}}}
    {\atom{ok} \sumExt \atom{no}}}
  \big) \\
  \pmove{} \hspace{4pt}
  & (s)\; \big(
  \sys {\pmv A} {\fact{x}{\overline{\atom{no}}}}
  \mid
  \sys {\pmv B} {\fact{y}{\atom{ok}}}
  \mid 
  \sys s {\bic
    {\overline{\atom{ok}} \sumInt \overline{\atom{no}}}
    {\atom{ok} \sumExt \atom{no}}}
  \big) \\
  \pmove{} \hspace{4pt}
  &  (s)\; \big(  
  \sys {\pmv A} {\pnil}
  \mid
  \sys {\pmv B} {\fact{y}{\atom{ok}}}
  \mid 
  \sys s {\gamma}
  \big)
\end{align*}
where $\gamma = \bic{\E}{\ready{\atom{no}}}$.
The system is then stuck, because
$\gamma$ is not allowing the \nrule{[Do]} step.
By Def.~\ref{def:culpable} we have 
$\csmiley{\pmv A}{\gamma}$, $\cfrown{\pmv B}{\gamma}$,
so {\pmv A} is honest in $S$ while {\pmv B} is not.
Actually, {\pmv B} has violated the contract agreed upon,
because he is waiting for a positive answer from the store, 
while in $c_{\pmv B}$ he also promised to accept a $\overline{\atom{no}}$.
By Def.~\ref{def:honest}, {\pmv B} is not honest, 
while we will show in \S~\ref{sect:realizability} that
{\pmv A} is honest (see Ex.~\ref{ex:store:realizes}).
\end{example}

We now define when a process enables a contract transition,
independently from the context. 
To do that, first we define the set $\readydo{s}{P}$ 
(after ``ready do''), which collects
all the atoms with an unguarded action $\fact{s}{\!}$ in $P$.

\begin{definition} \label{def:readydo}
For all $P$ and all $s$, we define the set of atoms
$\readydo{s}{P}$ as:
\[
  \readydo{s}{P} 
  \;\; = \;\;
  \setcomp
  {\atom{a}}
  {\exists \vec{u},P',Q,R\ .\ 
    P \equiv (\vec{u})\ (\fact{s}{\atom{a}}.P' + Q \mid R)
    \text{ and } s \not \in \vec{u}}
\]
\end{definition}

\noindent
Next, we check when a contract
``unblocks'' a set of atoms $\mathcal{X}$: 
e.g., if $\mathcal{X}$ accounts for at least one 
branch of an internal
choice, or for all the branches of an external choice.

\begin{definition} \label{def:unblocks}
For all sets of atoms $\mathcal{X}$ and for all $c \neq \cnil$,
we say that $\unblocks{c}{\mathcal{X}}$ iff:
\[
  \exists Y \in \rs{c}.\, Y \subseteq \mathcal{X} \cup \setenum{\atom{e}}
  \hspace{20pt} \text{or} \hspace{20pt}
  c = \ready{\atom{a}}.c' 
  \;\; \land \;\;
  \atom{a} \in \mathcal{X} \cup \setenum{\atom{e}}
\]
\end{definition}

\newcommand{\lemreadydo}{
For all $P$ 
and for all $\gamma = \bic{c}{d}$,
if $\unblocks{c}{\readydo{s}{P}}$ and
$S = (\vec{u}) ({\pmv A}[ P ] \mid s[\gamma] \mid S')$,
then either $\csmiley{\pmv A}{\gamma}$ or 
$S \pmove{\does[s]{\pmv A}{\atom{a}}}$.
}
\begin{lemma} \label{lem:readydo} 
\lemreadydo
\end{lemma}

The following theorem is the \coco counterpart 
of Theorem~\ref{lem:compliant-smiley}.
It states that, when a session $s$ is established 
between two participants {\pmv A} and {\pmv B}, 
{\pmv A} can always exculpate herself by performing 
(at most) two actions $\does{\pmv A}{-}$.
Note that when the contracts used to establish $s$ are compliant,
then we deduce the stronger thesis $\csmiley[s]{\pmv A}{S_{j}}$.

\newcommand{\lemdoubledo}{
Let $(S_i)_i$ be the following {\pmv A}-solo stable $\pmove{}$-trace,
with 
\(
  S_i = (\vec{u}_i)\;
  \big( 
  {\pmv A}[Q_i] \mid s[{\pmv A} \says c_i \mid {\pmv B} \says d_i] \mid S'_i
  \big)
\), 
and:
\[
  S_0
  \pmove{\mu_0}
  \cdots
  \pmove{\mu_{i-2}}
  S_{i-1}
  \pmove{\does[s]{\pmv A}{\atom{a}}} 
  S_{i}
  \pmove{\mu_{i}}
  \cdots
  \pmove{\mu_{j-2}}
  S_{j-1}
  \pmove{\does[s]{\pmv A}{\atom{b}}} 
  S_{j}
  \pmove{\mu_j}
  \cdots
\]
where $\mu_h \neq \does[s]{\pmv A}{-}$ for all $h \in [i,j-2]$.
Then,
either $c_j = \cnil$
or
$\csmiley[s]{\pmv A}{S_{j}}$.
}

\begin{theorem}[Factual exculpation] \label{lem:doubledo}
\lemdoubledo
\end{theorem}

The following theorem states the undecidability of honesty.

\newcommand{\thhonestyundecidable}{
The problem of deciding whether a participant $\sys {\pmv A}{P}$ 
is dishonest is recursively enumerable, but not recursive. 
}
\begin{theorem} \label{th:honesty-undecidable}
\thhonestyundecidable
\end{theorem}

\section{A criterion for honesty} \label{sect:realizability}

In this section we devise a sufficient criterion for honesty.
Actually, checking honesty is a challenging task
(indeed, by Th.~\ref{th:honesty-undecidable}, it is not even decidable),
because Def.~\ref{def:honest}
involves a universal quantification over \emph{all} possible contexts.
We will then provide a semantics of contracts and processes, 
that focusses on the actions performed by a single participant {\pmv A}, 
while abstracting from those made by the context.
Note that our abstract semantics assumes processes without top-level 
delimitations, in accordance with Def.~\ref{def:stable} 
which lifts such delimitations outside participants.
Further, we sometimes perform this lifting explicitly through the
$\open{-}$ operator.

\begin{definition}\label{def:abstract-contract}
For all participant names {\pmv A},
the abstract LTSs $\abscmove{}$ and $\abspmove[\pmv A]{}$ on
contracts and on processes, respectively, are defined by the rules in
Fig.~\ref{fig:abstract-co2},
where $\sigma: \vars \rightarrow \snames$.
\end{definition}

The intuition behind the abstract rules is provided by
Lemma~\ref{lem:abstract-contract} and Lemma~\ref{lem:co2-to-abstract} below,
which establish the soundness of the abstractions.

\newcommand{\lemabstractcontract}{
For all bilateral contracts $\gamma = \pmv A \says c \mid \pmv B \says d$:
\begin{enumerate}
\item \(
  \gamma
  \cmove{\pmv A \says \atom{a}}
  \pmv A \says c' \mid \pmv B \says d'
  \;\;\implies\;\;
  c \abscmove{\atom{a}} c' \;\land\; 
  (d \abscmove{ctx} d'
  \;\lor\;
  d \abscmove{\cnil} d')
\)

\item \(
  \gamma
  \cmove{\pmv A \says \atom{a}}
  \pmv A \says c' \mid \pmv B \says d'
  \;\land\;
  c \compliant d
  \;\;\implies\;\;
  c \abscmove{\atom{a}} c' \;\land\; d \abscmove{ctx} d'
\)
\end{enumerate}
}
\begin{lemma}\label{lem:abstract-contract}
\lemabstractcontract
\end{lemma}

Intuitively, a move of $\gamma$ is caused by an action performed by
one of its components $c$ and $d$. If $c$ moves, the $\abscmove{\atom
  a}$ rules account for its continuation. 
This might make $d$ commit to one of the branches of a sum, 
as shown in the $\abscmove{\ctx}$ rules.
Further, $c$ can perform an action not supported by $d$, by using a
$\nrule{[*Fail]}$ rule: accordingly, $\abscmove{\cnil}$ transforms $d$
into~$\cnil$. 
The compliance between $c$ and $d$ ensures the absence of such failure moves.

\begin{figure}[t]
\hrulefill
\[
\begin{array}{c}
    \sumI{\atom{a}}{c} \sumInt c'
    \abscmove{\atom{a}} c
    \hspace{20pt}
    \sumE{\atom{a}}{c} \sumExt c'
    \abscmove{\atom{a}} c
    \hspace{20pt}
    \ready{\atom{a}}.\; c \abscmove{\atom{a}} c 
    \hspace{20pt}
    \sumI{\atom{a}}{c} \sumInt c'
    \abscmove{\atom{a}} E
    \hspace{20pt}
    \sumE{\atom{a}}{c} \sumExt c'
    \abscmove{\atom{a}} E
\\[10pt]
    \SumInt{\atom{a}_i}{c_i} \abscmove{\cnil} \cnil        
    \hspace{14pt}
    \SumExt{\atom{a}_i}{c_i} \abscmove{\cnil} \cnil        
    \hspace{14pt}
    \SumExt{\atom{a}_i}{c_i} \abscmove{ctx} \ready{\atom{a}_n} . \, c_n 
    \hspace{14pt}
    \sumI{\atom{a}}{c}       \abscmove{ctx} \ready{\atom{a}}  . \, c 
    \hspace{14pt}
    c \abscmove{\ctx} c
\\[10pt]
  \pi.P + Q \mid R \abspmove{\pi} \begin{cases}
      \open{ \freeze x {\pmv A \says c} \mid P \mid R }
      & \!\!\!\text{if } \pi = \tell{\pmv A}{\freeze x c} \\[5pt]
      \open{P \mid R}\sigma 
      & \!\!\!\text{otherwise}
    \end{cases}
  \hspace{15pt}
  \begin{array}{l}
  P \abspmove{\ctx} \ \freeze{x}{\pmv B \says c} \mid P
  \;\mbox{ if $\pmv B \neq \pmv A$}
  \\[5pt]
  P \abspmove{\ctx} P \sigma
  \end{array}
\\[15pt]
\open{P} = P' \mbox{ where $P \equiv (\vec{u}_i) P'$ and no delimitation of
$P'$ can be brought to the top level}
\end{array}
\]
\hrulefill
\vspace{-5pt}
\caption{Abstract LTSs for contracts and processes 
($\sigma: \vars \rightarrow \snames$, 
name $\pmv A$ in $\abspmove[\pmv A]{}$ is omitted).}
\label{fig:abstract-co2}
\vspace{-10pt}
\end{figure}

\newcommand{\lemcocotoabstract}[0]{
For each (finite or infinite) stable $\pmove{}$-trace $(S_i)_i$, 
with $S_i = (\vec{u}_i) (\sys{\pmv A}{Q_i} \ |\ S_i')$,
there exists a $\abspmove{}$-trace
$Q_0 \abspmove{\mu_0} Q_1 \abspmove{\mu_1} Q_2 \abspmove{\mu_2} \cdots$
where $\mu_i = \pi$ if $S_i \pmove{\pmv A \says \pi} S_{i+1}$, and
$\mu_i = \ctx$ otherwise.
Moreover, if $(S_i)_i$ is fair, then $(Q_i)_i$ is 
$\setenum{\unblocked,\tell{}}$-fair.
}
\begin{lemma}\label{lem:co2-to-abstract}
\lemcocotoabstract
\end{lemma}

In the above lemma, each step of the whole system
might be due to either the process $Q_i$ or its
context. 
If $Q_i$ fires a prefix $\pi$, then it changes according to
the $\abspmove{\pi}$ rule in Fig.~\ref{fig:abstract-co2}. 
In particular,
that accounts for $\tell{\pmv A}{-}$ adding further latent contracts to
$Q_i$, as well as $\fuse[]{}$ possibly instantiating variables. 
Newly exposed delimitations are removed using $\open{-}$:
indeed, they already appear in $\vec{u}_i$, since the trace is stable.

We now define when a process $P$ ``$\sharp$-realizes'' 
a contract $c$ in a session $s$ (written $\realizes{P}{c}{s}$), 
without making any assumptions about its context. 
Intuitively, $\realizes{P}{c}{s}$ 
holds when 
$(1)$ $P$ eventually enables the $\fact{s}{}$ actions mandated by $c$, and 
$(2)$ in the abstract LTS $\abspmove{}$,
the continuation of $P$ after firing some $\fact{s}{}$ 
must realize the continuation of $c$ (under $\abscmove{}$).
Note that $P$ is not required to actually perform the
relevant $\fact{s}{}$, because the context might prevent $P$ from doing so.
For instance, in the system
$\sys{\pmv A}{P} \mid  \sys{s}{\bic{c}{\ready{\atom{a}}.d}}$
the process $P$ can not fire any $\fact{s}{\!}$.

\begin{definition}\label{def:realizes}
Given a session $s$ and a participant {\pmv A}, 
we define the relation $\realizes[\pmv A]{}{}{s}$  
\mbox{(``$\sharp$-realizes'')}
between processes and contracts 
as the largest relation such that, 
whenever $\realizes[\pmv A]{P_0}{c}{s}$, then
for each $\setenum{\unblocked,\tell{}}$-fair 
$\abspmove[\pmv A]{}$-trace $(P_i)_i$
without labels $\fact{s}{-}$, we have:
\begin{enumerate}

\item\label{def:realizes:1}
 $\exists k .\; \forall i \geq k .\; \unblocks{c}{\readydo{s}{P_i}}$

\item\label{def:realizes:2}
 $\forall i,\atom{a},P',c' .\; \big( 
  P_i \abspmove{\fact{s}{\atom{a}}} P' 
  \;\land\; 
  c \abscmove{\atom{a}} c' 
  \implies \realizes[\pmv A]{P'}{c'}{s}
  \big)$
\end{enumerate}
\end{definition}

\begin{example} \label{ex:store:realizes}
  Recall the online store {\pmv A} from Ex.~\ref{ex:store:co2}.
  We show that $\realizes{X\setenum{\bind{x}{s}}}{c_{\pmv A}}{s}$.
  First note that transitions in
  $\setenum{\unblocked,\tell{}}$-fair $\abspmove{}$-traces without
  $\fact s {}$ from $X\setenum{\bind x s}$ can only be labelled with
  $\ctx$.
  Thus, each process $P_i$ on such traces has the form
  $X\setenum{\bind x s} \mid K_i$, for some $K_i$.
  We have $\readydo s {P_i} = \readydo s {X\setenum{\bind x s}} = \{\atom{addToCart},
  \atom{creditCard} \}$. 
  Moreover,
  $\unblocks{c_{\pmv A}}{\readydo s {X\setenum{\bind x s}}}$ 
  hence condition~\eqref{def:realizes:1} of
  Def.~\ref{def:realizes} holds.
  For condition~\eqref{def:realizes:2}, if
  \(
  c_{\pmv A} \abscmove{\atom{creditCard}} c' = \overline{\atom{accept}} \sumInt
  \overline{\atom{reject}}
  \)
  and
  \(
  P_i \abspmove{\fact s
    {\atom{creditCard}}} P' = \tau.\fact{s}{\overline{\atom{accept}}}
  + \tau.\fact{s}{\overline{\atom{reject}}} \mid K_i
  \)
  then $\realizes{P'}{c'}{s}$.
  Actually, all processes on a
  $\setenum{\unblocked,\tell{}}$-fair $\abspmove{}$-traces without
  $\fact s {}$ from $P'$ have either the form
  $\fact{s}{\overline{\atom{accept}}} \mid K$ or the form
  $\fact{s}{\overline{\atom{reject}}} \mid K$.
  For the recursive case, $c_{\pmv A} \abscmove{\atom{addToCart}} c_{\pmv A}$ and $P_i
  \abspmove{\fact s {\atom{addToCart}}} X\setenum{\bind x s}$, 
  hence $\realizes {X\setenum{\bind x s}} {c_{\pmv A}} s$ by coinduction.
  Note that the case $c_{\pmv A} \abscmove{\atom e}$ did not apply, 
  because $P_i$ cannot take $\abspmove{}$-transitions labelled 
  $\fact{s}{\atom e}$.
\end{example}

Theorem~\ref{lem:realizes-step-co2} below 
establishes an invariant of system transitions. 
If a participant $\sys{\pmv A}{Q_0}$ $\sharp$-realizes a stipulated contract
$c_0$, then in each evolution of the system the descendant of
$\sys{\pmv A}{Q_0}$ still $\sharp$-realizes the related descendant of $c_0$.
The theorem only assumes that $c_0$ is in a session with a compliant
contract, as it is the case after firing a $\fuse{}$.

\newcommand{\lemrealizesstepcoco}{
Let $\!(S_i)_i$ be a stable $\!\pmove{}$-trace with
\(
  S_i = 
  (\vec{u}_i)
  ( 
  {\pmv A}[Q_i] \!\mid\! s[\bic{c_i\!}{d_i}] \!\mid\! S_i'
  )
\)
for all $i$.
If $c_0 \compliant d_0$ and $\realizes[\pmv A]{Q_0}{c_0}{s}$,
then $\realizes[\pmv A]{Q_i}{c_i}{s}$ for all $i$.
}
\begin{theorem}\label{lem:realizes-step-co2}
\lemrealizesstepcoco
\end{theorem}

We now define when a participant is \emph{\canonical}.
Intuitively, we classify as such a participant ${\pmv A}[P]$ when, 
for all prefixes $\tell{}{\freeze{x}{c}}$ contained in $P$,
the continuation $Q$ of the prefix 
$\sharp$-realizes $c$.
We also require that the session variable $x$ cannot be used 
by any process in parallel with $Q$,
because such processes could potentially compromise the ability
of $Q$ to realise $c$ (see Ex.~\ref{ex:store:xsafe}).
\begin{definition}[\canonical\ participant]\label{def:canonical}
A participant $\sys {\pmv A} P$ is \emph{\canonical} iff
$P$ does not contain $\freeze{y}{{\pmv A} \says c}$, and for all
linear contexts $\mathcal C(\bullet)$, $x$, $c$, $Q$, $R$, and $s$ fresh in $P$
\[
  P = \mathcal{C}(\tell{}{\freeze{x}{c}}.Q + R) 
  \;\;\implies\;\;
  \realizes[\pmv A]{\open{Q\setenum{\bind{x}{s}}}}{c}{s} 
  \;\;\land\;\; 
  \mathcal{C} \textit{ is } x\textit{-safe}
\]
where $\mathcal C(\bullet)$  is $x$-safe iff
$\exists \mathcal{C}'.\; \mathcal{C}(\bullet) = \mathcal{C}'((x) \bullet)$
or $\mathcal{C}$ is free from $\fact{x}{-}$.
\end{definition}

\begin{example} \label{ex:store:xsafe}
Substitute $Q = \fuse{x}{}.\fact{x}{\atom{creditCard}}$ for
$\fuse{x}{}$ in the process $P_{\pmv A}$ from Ex.~\ref{ex:store:co2}.
Then $\sys{\pmv A}{P_{A}}$ is not honest, because
{\pmv A} cannot complete her contract
if the $\fact{x}{}$ within $Q$ is performed.
However, the modified $\sys{\pmv A}{P_{A}}$ violates $x$-safety,
hence it is not \canonical.
\end{example}

The following lemma relates \canonicity\ with the abstract semantics 
of processes.
If a \canonical\ process $P$ abstractly fires a $\tell{}{\freeze{x}{c}}$, then 
the continuation of $P$ realises $c$ (item~1).
Also, \canonicity\ is preserved under abstract transitions (item 2).

\newcommand{\lemcanonicalstep}{
For all \canonical\ participants $\sys {\pmv A} P$, such that $P=\open{P}$:
\begin{enumerate}

\item if $P \abspmove[]{\tell{\pmv B}{\freeze{x}{c}}} P'$, then
$\realizes[\pmv A]{P'\setenum{\bind{x}{s}}}{c}{s}$, for all $s$ fresh in $P$.

\item if $P \abspmove[]{} P'$, then $\sys {\pmv A} {P'}$ is \canonical.

\end{enumerate}
}
\vbox{
\begin{lemma} \label{lem:canonical-step}
\lemcanonicalstep
\end{lemma}}

Our main result states that \canonicity\ suffices to ensure honesty.
Note that while honesty, by Def.~\ref{def:honest}, considers all the
(infinite) possible contexts, \canonicity\ does not.  Hence, while
verifying honesty can be unfeasible in the general case, it can also
be ensured by establishing \canonicity, which is more amenable to
verification.  
For instance, for finite control processes~\cite{dam97} it is possible
to decide \canonicity\ e.g. through model-checking. In fact, in these
processes parallel composition cannot appear under recursion, hence
their behaviour can be represented with finitely many states.

\newcommand{\thcanonicalissaint}{
All \canonical\ participants are honest.
}
\begin{theorem} \label{th:canonical-is-saint}
\thcanonicalissaint
\end{theorem}

Noteworthily, by Theorem~\ref{th:canonical-is-saint} we can establish
that all the participants named {\pmv A} in 
Examples~\ref{ex:store:co2},~\ref{ex:voucher:co2}, and~\ref{ex:travel:co2}
are honest.
This is obtained by reasoning as in Example~\ref{ex:store:realizes}.
Instead, participant {\pmv A} in Example~\ref{ex:choice-via-ask}
is honest but not \canonical.

\section{Related Work and Conclusions} \label{related-work}

We have developed a formal model for reasoning about 
contract-oriented systems.
Our approach departs from the common principle that contracts are always respected
after they are agreed upon. 
We represent instead the more realistic situation
where promises are not always kept.
The process calculus \coco\!~\cite{BTZ11ice} 
allows participants to advertise contracts, 
to establish sessions with other participants with compliant contracts,
and to fulfill them (or choose not to).
Remarkably, instead of defining an ad-hoc contract model,
we have embedded the contract theory of~\cite{Castagna09toplas} within \coco\!.
To do that, we have slightly adapted the contracts
of~\cite{Castagna09toplas} in order to define culpability,
and we have specialized \coco\ accordingly at the system-level.
The main technical contribution of this paper is a criterion for
deciding when a participant always respects the advertised contracts 
in all possible contexts.
This is not a trivial task, especially when multiple sessions are
needed for realizing a contract (see e.g.\ Ex.~\ref{ex:voucher:co2}
and~\ref{ex:travel:co2}) or when participants want to inspect the
state of a contract to decide how to proceed next (see e.g.\
Ex.~\ref{ex:choice-via-ask}).

At the best of our knowledge, this is the first paper
that addresses the problem of
establishing when a participant is honest
in a contract-based system populated by
dishonest participants.
Several papers investigated the use of contracts in
concurrent systems; however, they typically focus on
coupling processes which statically guarantee
conformance to their contracts.
This is achieved e.g.\
by typing~\cite{Bocchi10concur,Carpineti06basic,Castagna09toplas},
by contract-based process synthesis~\cite{Buscemi11tgc},
or by approaches based on behavioural preorders~\cite{Bravetti07sc}.

The process calculus \coco has been introduced in~\cite{BTZ11ice} as a
generic framework for relating different contract models; the variant
in this paper has been obtained by instantiating it with the
contracts of~\cite{Castagna09toplas}.
Some primitives, e.g.\ multiparty $\fuse{}{}$, 
have been consequently simplified.
In~\cite{BTZ11ice}, a participant {\pmv A} is honest when
{\pmv A} becomes not culpable from a certain execution step;
here, we only require that, whenever {\pmv A} is culpable,
then she can exculpate herself by performing some actions.
This change reflects the fact that bilateral contracts
\`a la~\cite{Castagna09toplas} can describe endless
interactions.
The notion of compliance in~\cite{Castagna09toplas} is asymmetric.
Namely, if $c$ is the client contract and $d$ is the server contract,
then $c$ and $d$ are compliant if $c$ always reaches a success state
or engages $d$ in an endless interaction.
In our model instead compliance is symmetric: the server contract,
too, has to agree on when a state is successful.
The LTS semantics of unilateral contracts in~\cite{Castagna09toplas} yields identical synchronization trees for internal and external choice;
to differentiate them, one has to consider their ready sets.
We instead give semantics to {\em bilateral} contracts,
and distinguish between choices at the LTS level.
Note that we do not allow for unguarded sums, unlike~\cite{Castagna09toplas}.
Were these be allowed, we would have to deal e.g.\ with
a participant {\pmv A} with a contract of the form
$\sumI{\atom{a}}{c_0} \sumInt (\sumE{\atom{b}}{c_1} \sumExt \sumE{\atom{c}}{c_2})$.
According to our intuition {\pmv A} should be culpable, because of the internal choice.
If {\pmv A} legitimately chooses not to perform $\atom{a}$,
to exculpate herself she would have to wait for the other participant
to choose (internally) between $\atom{b}$ and $\atom{c}$.
Therefore, {\pmv A} can exculpate herself only if the other participant permits her to.
By contrast, by restricting to guarded sums our theory enjoys the nice feature that
a culpable participant can always exculpate herself by
performing some actions, which “pass the buck” to the other participant
(Theorems~\ref{lem:compliant-smiley} and~\ref{lem:doubledo}).

Design-by-contract is transferred in~\cite{Bocchi10concur} to distributed
interactions modelled as (multiparty) asserted global types.
The projection of asserted global types on local ones allows for
the automatic generation of monitors whereby incoming messages are
checked against the local contract.
Such monitors have a ``local'' view of the computation, i.e.\ they
can detect a violation but cannot, in general, single out the
culpable component.
In fact, a monitor cannot know if an expected message is not
delivered because the partner is violating his contract, or because
he is blocked on interactions with other participants.
Conversely, our notion of \emph{honesty} singles out
culpable components during the computation.
An interesting problem would be to investigate how our notion of
culpability could be attained within the approach in~\cite{Bocchi10concur}.
In fact, this seems to be a non trivial problem, even if forbidding
communication channels shared among more than two participants.

Contracts are rendered in~\cite{Buscemi07ccpi,Buscemi11tgc}
as soft constraints (values in a c-semiring) 
that allow for different levels of agreement between contracts.
When matching a client with a service, the constraints are composed.
This restricts the possible interactions to those acceptable (if any)
to both parties.
A technique is proposed in~\cite{Buscemi11tgc}
for compiling clients and services so
that, after matching, both actually behave according to the mutually
acceptable interactions, and reach success without getting stuck.
Our framework is focused instead on blaming participants, and on
checking when a participant is honest, i.e.\ always able to avoid
blame in all possible contexts.
The use of soft constraints in a context where participants can be
dishonest seems viable, e.g.\ by instantiating the abstract contract
model of \coco with the contracts in~\cite{Buscemi11tgc}.  
A challenging task would be that of defining culpability in such setting.

\paragraph{Acknowledgments.}
This work has been partially supported by 
by Aut.\ Region of Sardinia under grants L.R.7/2007 CRP2-120 (Project TESLA) 
and CRP-17285 (Project TRICS),
and by the Leverhulme Trust Programme Award ``Tracing Networks''.

\bibliographystyle{abbrv}
\bibliography{main}

\appendix
\newpage
\section{Proofs for Section~\ref{sect:contracts}}
\label{proofs-contracts}
\label{proofs-culpability}

The following lemma ensures that transition steps preserve the invariant
required in Definition~\ref{def:contracts:syntax},
i.e.\ that only one $\ready{}\!$ can occur in a bilateral contract.

\newcommand{\lemnodoubleready}{
For all $\gamma$,
if $\gamma \cmove{} {\pmv A} \says \ready{\atom{a}}.c \mid {\pmv B} \says d$, then
$d$ is $\ready{}\!$-free.
}
\begin{applemma} \label{lem:no-double-ready}
\lemnodoubleready
\end{applemma}
\begin{proof}
By Definition~\ref{def:contracts:semantics}, a $\ready{}\!$ can only occur
at top-level of a contract.
By Definition~\ref{def:contracts:syntax}, only one $\ready{}\!$ can occur
in $\gamma$.
The thesis then follows by straightforward analysis of the rules in 
Figure~\ref{fig:contracts:semantics:nonfail}.
\qed
\end{proof}

The following lemma states that bilateral contracts evolve deterministically
under the actions performed by participants. 
This agrees with the intuition that, 
in a contract involving the participants {\pmv A} and {\pmv B}
(and no other third parties)
the duties of {\pmv A} and {\pmv B} only depend on the choices
performed by  {\pmv A} and {\pmv B}, and not on some external entity.
Notice that, when {\pmv A} and {\pmv B} advertise two internal choices
and evolve through \nrule{[IntInt]}, 
determinism is ensured by the fact that one of the choices is a singleton.
Otherwise, {\pmv A} and {\pmv B} could either succeed by internally choosing 
the same action, or fail by choosing different ones.
Let e.g.\
\(
  \gamma =
  {\pmv A} \says \sumI{\atom{a}}{c_1} \sumInt \sumI{\atom{b}}{c_2} \mid
  {\pmv B} \says \sumI{\bar{\atom{a}}}{d_1} \sumInt \sumI{\bar{\atom{b}}}{d_2}  
\).
Were \nrule{[IntInt]} allowing $\gamma$ to evolve to 
${\pmv A} \says c_1 \mid {\pmv B} \says d_1$
with label ${\pmv A} \says \atom{a}$,
then we would lose determinism, since rule \nrule{[IntIntFail]}
allows $\gamma$ to also evolve to 
${\pmv A} \says c_1 \mid {\pmv B} \says \cnil$.
Note however that determinism would still hold for compliant contracts.

\begin{barttheorem}{Lemma~\ref{lem:stuck-iff-nil}}
\lemstuckiffnil
\end{barttheorem}
\begin{proof}
Let $\gamma = {\pmv A} \says c \mid {\pmv B} \says d$.

For the ``only if'' part, if $c = d = \cnil$ then no rules 
in Definition~\ref{def:contracts:semantics} can be applied.
Therefore, $\gamma$ is stuck.

For the ``if'' part, assume by contradiction that $c \neq \cnil$
(the case $d \neq \cnil$ is symmetric, so we omit it).
We have the following exhaustive cases:
\begin{itemize}

\item if $c = \SumInt{\atom{a}_i}{c_i}$, then
then $\gamma$ can take a transition through one of the rules
$[\nrule{IntExt}]$, $[\nrule{IntInt}]$, $[\nrule{IntExtFail}]$, $[\nrule{IntIntFail}]$.

\item if $c = \SumExt{\atom{a}_i}{c_i}$, then
then $\gamma$ can take a transition through one of the rules 
$[\nrule{IntExt}]$, $[\nrule{ExtExt}]$, $[\nrule{IntExtFail}]$, $[\nrule{ExtExtFail}]$.

\item if $c = \ready{\atom{a}} . c'$, 
then $\gamma$ can take a transition through rule $[\nrule{Rdy}]$

\end{itemize}
In each case, we have proved that $\gamma$ can take a transition;
therefore, $\gamma$ is not stuck.
\qed
\end{proof}

\begin{applemma} \label{lem:rs-nonempty}
For all contracts $c$, $\rs{c} \neq \emptyset$.
\end{applemma}
\begin{proof}
Straightforward case analysis of Def.~\ref{def:rs}.
\qed
\end{proof}

\begin{barttheorem}{Lemma~\ref{lem:contracts:determinism}}
\lemcontractsdeterminism
\end{barttheorem}
\begin{proof}
Let $\gamma = {\pmv A} \says c \mid {\pmv B} \says d$,
and w.l.o.g.\ assume that $\mu = {\pmv A} \says \atom{a}$.
According to the structure of $c$ and $d$, and to
the rules in Figures~\ref{fig:contracts:semantics:nonfail}
and~\ref{fig:contracts:semantics:fail},
each rule is able to generate at most one $\mu$ transition
for $\gamma$. It is therefore enough to consider the set of 
applicable rules. We have the following exhaustive, non-overlapping cases:
\begin{enumerate}
\item $c$ internal sum, $d$ external sum 
$\Longrightarrow$ rules \nrule{[IntExt]}, \nrule{[IntExtFail]}
\item $c$ internal sum, $d$ internal sum
$\Longrightarrow$ rules \nrule{[IntInt]}, \nrule{[IntIntFail]}
\item $c$ external sum, $d$ internal sum
$\Longrightarrow$ symmetric of rules \nrule{[IntExt]}, \nrule{[IntExtFail]}
\item $c$ external sum, $d$ external sum
$\Longrightarrow$ rules \nrule{[ExtExt]}, \nrule{[ExtExtFail]}
\item $c$ is a $\ready{}\!$
$\Longrightarrow$ rule \nrule{[Rdy]}.
\end{enumerate}
Note that we are not considering the case where $d$ is a $\ready{}\!$ and
$c$ is not, because in such case {\pmv A} cannot perform any action.
We now show that, for all $\nrule{x} \in \setenum{\nrule{IntExt,IntInt,ExtExt}}$,
the rules $[\nrule{x}]$ and $[\nrule{xFail}]$ are mutually exclusive.
We have three cases:
\begin{itemize}

\item $[\nrule{IntExt}]$.
Let $c = \SumInt[i \in I]{\atom{a}_i}{c_i}$, and
let $d = \SumExt[j \in J]{\atom{b}_j}{d_j}$
(symmetric case is similar).
If rule $[\nrule{IntExt}]$ can be applied, then
\(
  \exists i \in I, j \in J.\;
  \atom{a} = \atom{a}_i \text{ and } \atom{b}_j = \bar{\atom{a}}
\),
which makes false the precondition of \nrule{[IntExtFail]}.
Conversely, if \nrule{[IntExtFail]} can be applied, 
then $\atom{a} \not\in \co{\setenum{\atom{b}_j}_{j \in J}}$,
and so \nrule{[IntExt]} cannot be applied.

\item $[\nrule{IntInt}]$.
Let $c = \SumInt[i \in I]{\atom{a}_i}{c_i}$, and
let $d = \SumInt[j \in J]{\atom{b}_j}{d_j}$.
If rule $[\nrule{IntInt}]$ can applied, then
\(
  \exists i \in I.\; \atom{a} = \atom{a}_i \text{ and } 
  d = \sumI{\bar{\atom{a}}}{d'}
\)
for some $d'$, 
which makes false the precondition of \nrule{[IntIntFail]}.
Conversely, if \nrule{[IntIntFail]} can be applied, 
then $\co{\setenum{\atom{b}_j}_{j \in J}} \neq \setenum{\atom{a}}$, 
which prevents from using the rule \nrule{[IntInt]}.

\item $[\nrule{ExtExt}]$.
Let $c = \SumExt[i \in I]{\atom{a}_i}{c_i}$, and
let $d = \SumExt[j \in J]{\atom{b}_j}{d_j}$.
If rule $[\nrule{ExtExt}]$ can applied, then
\(
  \exists i \in I, j \in J.\;
  \atom{a} = \atom{a}_i \text{ and } \atom{b}_j = \bar{\atom{a}}
\),
and so the the precondition of \nrule{[ExtExtFail]} is false.
Conversely, if the precondition of \nrule{[ExtExtFail]} is true,
then there exist no $i,j$ such that 
$\atom{a}_i = \bar{\atom{b}}_j$,
and so \nrule{[ExtExt]} cannot be applied.
\qed
\end{itemize}
\end{proof}


\newcommand{\lemcompliantnil}[0]
{
For all contracts $c,d$, 
if $c \compliant d$ then $c \neq \cnil$ and $d \neq \cnil$.
}
\begin{applemma} \label{lem:compliant-nil}
\lemcompliantnil
\end{applemma}
\begin{proof}
By contradiction, assume w.l.o.g.\ that $d = \cnil$.
By Def.~\ref{def:rs}, $\rs{d} = \setenum{\emptyset}$.
Therefore, by condition $(1)$ of Def.~\ref{def:compliance}, 
for all $\mathcal{Y} \in \rs{c}$ it must be $\ready{}\! \in \mathcal{Y}$.
By Lemma~\ref{lem:rs-nonempty}, $\rs{c} \neq \emptyset$,
and so by Def.~\ref{def:rs} and by the fact that $\ready{}\!$ can only
occur at top-level in a contract, it must be $c = \ready{\atom{a}}. c'$,
for some $\ready{\!}$-free $c'$.
By the rule $[\nrule{Rdy}]$ in
Fig.~\ref{fig:contracts:semantics:nonfail}, 
it follows that
${\pmv A} \says c \mid {\pmv B} \says \cnil \cmove{{\pmv A} \says
\atom{a}} {\pmv A} \says c' \mid {\pmv B} \says \cnil$.
By condition $(2)$ in Def.~\ref{def:compliance}, it should be 
$c' \compliant \cnil$.
The whole argument used above can be replayed to deduce that $c'$ must be
of the form $\ready{\atom{a}}. c''$ --- contradiction, because
$c'$ is without $\ready{}\!$. 
\qed
\end{proof}

\begin{barttheorem}{Lemma~\ref{lem:compliant-fail}}
\lemcompliantfail
\end{barttheorem}
\begin{proof}
For the ``only if'' part, assume that $c \compliant d$.
Assume that
$\gamma \cmove{}^n {\pmv A} \says c' \mid {\pmv B} \says d'$.
We proceed by induction on $n$.
For the base case $n = 0$,
by Lemma~\ref{lem:compliant-nil} it follows that
$c \neq \cnil$ and $d \neq \cnil$.
For the inductive case, 
let $\gamma \cmove{} {\pmv A} \says c'' \mid {\pmv B} \says d''$.
By the condition $(2)$ of Def.~\ref{def:compliance}, $c'' \compliant d''$,
and so by the induction hypothesis we conclude. 

For the ``if'' part, assume that all the descendants of 
$\gamma = {\pmv A} \says c \mid {\pmv B} \says d$ 
have non-$\cnil$ contracts.
Let $\mathcal{R}$ be the following relation on contracts:
\[
  c' \mathcal{R} d' 
  \quad \text{ iff } \quad
  \gamma  \cmove{}^* {\pmv A} \says c' \mid {\pmv B} \says d'
\]
We will prove that $\mathcal{R}$ satisfies both the conditions $(1)$
and $(2)$ of Def.~\ref{def:compliance}.
Since $\compliant$ is the largest relation satisfying these conditions,
we shall then conclude that $c \compliant d$.
The condition $(2)$ holds by construction.
For the condition $(1)$, let $c' \mathcal{R} d'$, 
and assume by contradiction that: 
\begin{equation} \label{eq:compliant-fail}
  \exists \mathcal{X} \in \rs{c'},\ \mathcal{Y} \in \rs{d'} .\;\;
  \co{\mathcal{X}} \cap \mathcal{Y} = \emptyset \text{ and }
  \ready{} \not\in (\mathcal{X} \cup \mathcal{Y}) \setminus (\mathcal{X} \cap \mathcal{Y})
\end{equation}
We now proceed by cases on the syntax of $c'$ and $d'$.
Note that $\ready{\!}$ may occur at most once in 
$\bic{c'}{d'}$ because of the syntactic
restriction on bilateral contracts (which is preserved by transitions,
as stated by Lemma~\ref{lem:no-double-ready}).
Hence $\ready{} \not \in \mathcal{X} \cap \mathcal{Y}$, 
which with \eqref{eq:compliant-fail}
actually implies $\ready{} \not \in \mathcal{X} \cup \mathcal{Y}$, proving that
$\ready{}\!$ does not occur at all in $c'$ nor in $d'$.
Then, there are the following exhaustive cases (symmetric cases are omitted):
\begin{itemize}

\item if $c' = \SumInt{\atom{a}_i}{c'_i}$ and 
$d' = \SumExt[i \in J]{\atom{b}_i}{d'_i}$, then
by Def.~\ref{def:rs} and~\eqref{eq:compliant-fail}
there exist $\mathcal{X} = \setenum{\atom{a}_i}$ with 
$\atom{a}_i \not\in \co{\setcomp{\atom{b}_j}{j \in J}}$.
Then, by the rule $[\nrule{IntExtFail}]$, it follows that
$\gamma \cmove{{\pmv A} \says \atom{a}_i} \bic{\E}{\cnil}$
--- contradiction.

\item if $c' = \SumInt{\atom{a}_i}{c'_i}$ and 
$d' = \SumInt[j \in J]{\atom{b}_i}{d'_i}$, then
by Def.~\ref{def:rs} and~\eqref{eq:compliant-fail}
there exist $\mathcal{X} = \setenum{\atom{a}_i}$ and $\mathcal{Y} = \setenum{\atom{b}_j}$
such that $\atom{a}_i \neq \bar{\atom{b}}_j$.
Hence $\setenum{\atom{a}_i} \neq \co{\setcomp{\atom{b}_j}{j \in J}}$.
Then, by the rule $[\nrule{IntIntFail}]$, it follows that
$\gamma \cmove{{\pmv A} \says \atom{a}_i} \bic{\E}{\cnil}$
--- contradiction.

\item if $c' = \SumExt[i \in I]{\atom{a}_i}{c'_i}$ and 
$d' = \SumExt[j \in J]{\atom{b}_j}{d'_j}$, then
by Def.~\ref{def:rs} and~\eqref{eq:compliant-fail},
$\co{\setcomp{\atom{a}_i}{i \in I}} \cap \setcomp{\atom{b}_j}{j \in J}
= \emptyset$.
Then, by the rule $[\nrule{ExtExtFail}]$, it follows that
$\gamma \cmove{{\pmv A} \says \atom{a}_i} \bic{\E}{\cnil}$
--- contradiction.
\qed
\end{itemize}
\end{proof}

\begin{appdefinition}[Dual contract] \label{def:dual}
For all $\ready{}\!$-free contracts $c$,
let the contract $\dual{c}$ be inductively defined as follows:
\begin{align*}
\dual{\SumInt[i \in I]{\atom{a}_i}{c_i}} & =
\SumExt[i \in I]{\bar{\atom{a}_i}}{\dual{c_i}} 
&& \dual{\rec{X}{c}} = \rec{X}{\dual{c}} \\
\dual{\SumExt[i \in I]{\atom{a}_i}{c_i}} & =
\SumInt[i \in I]{\bar{\atom{a}_i}}{\dual{c_i}} 
&& \dual{X} = X
\end{align*}
\end{appdefinition}

\begin{applemma} \label{lem:compliant-dual}
For all $\cnil$-free and $\ready{}\!$-free $c$, $c \compliant \dual{c}$.
\end{applemma}
\begin{proof}
We will prove the following three properties,
which hold for all $\cnil$-free and $\ready{}\!$-free contracts $c$:
\begin{subequations}
\begin{align}
\label{eq:dual:1}
& \bic{c}{\dual{c}} \cmove{} \bic{c'}{d'}
\\
\nonumber
& \implies
\exists \atom{a},f \text{ without } \cnil :
c' \equiv f \text{ and } d' \equiv
\ready{\atom{a}}.\dual{f} 
\\
\nonumber
& \hspace{80pt} \text{ or } 
c' \equiv \ready{\atom{a}}.\dual{f} \text{ and } d' \equiv f
\\
\label{eq:dual:2}
& \bic{c}{\ready{\atom{a}}.\dual{c}} \cmove{} 
\gamma' 
\implies
\gamma' = \bic{c}{\dual{c}} \\
\label{eq:dual:3}
& \bic{\ready{\atom{a}}.c}{\dual{c}} \cmove{} 
\gamma'
\implies
\gamma' = \bic{c}{\dual{c}}
\end{align}
\end{subequations}

For~\eqref{eq:dual:1}, by Def.~\ref{def:dual} we have that
$c$ is an internal sum and $\dual{c}$ is an external sum, or
\emph{vice versa}.
W.l.o.g.\ assume that $c = \SumInt[i \in I]{\atom{a}_i}{c_i}$
(the case of an external sum is similar).
Note that the \nrule{[IntExtFail]} cannot be applied, since its precondition
$\exists i \in I .\ \forall j \in I.\ \atom{a}_i \neq \bar{\atom{a}_j}$ is false.
Therefore, the only applicable rule is \nrule{[IntExt]}, which
gives the desired conclusion.
Properties~\eqref{eq:dual:2} and~\eqref{eq:dual:3} hold trivially
by Def.~\ref{def:contracts:semantics}. 
Note that, under their hypotheses, the only applicable rule  is \nrule{[Rdy]}.

Taken together,~\eqref{eq:dual:1},~\eqref{eq:dual:2} and~\eqref{eq:dual:3}
guarantee that, 
if $\gamma = {\pmv A} \says c \mid {\pmv B} \says \dual{c}
\cmove{}^* {\pmv A} \says c' \mid {\pmv B} \says d'$,
then $c' \neq \cnil$ and $d' \neq \cnil$. 
By Lemma~\ref{lem:compliant-fail}, this enables us to conclude that 
$c \compliant \dual{c}$.
\qed
\end{proof}

\begin{barttheorem}{Lemma~\ref{lem:compliant-exists}}
\lemcompliantexists
\end{barttheorem}
\begin{proof}
If $c$ is $\ready{}\!$-free, then the thesis immediately follows from 
Lemma~\ref{lem:compliant-dual}, by choosing $d = \dual{c}$.
If $c = \ready{\atom{a}}.c'$, then by Definition~\ref{def:contracts:syntax}
$c'$ must be $\ready{}\!$-free.
Therefore, by Lemma~\ref{lem:compliant-dual}
$c' \compliant \dual{c'}$.
Let $d = \dual{c'}$.
The item $(1)$ of Definition~\ref{def:compliance} holds, because
$\rs{c} = \setenum{\setenum{\ready{\!}}}$.
The item $(2)$ also holds, because
there exists a unique transition from ${\pmv A} \says c \mid {\pmv B} \says d$,
leading to ${\pmv A} \says c' \mid {\pmv B} \says d$,
by the rule \nrule{[Rdy]},
and we have that $c' \compliant d$.
\qed
\end{proof}

\begin{applemma} \label{lem:compliant-inversion}
For all contracts $c,d$, if $c \compliant d$ then:
\begin{subequations}
\begin{align}
\label{eq:compliant-inversion:int}
  c = \SumInt[i \in I]{\atom{a}_i}{c_i} 
  \quad
  & \implies
  && d  = \SumExt[i \in J]{\bar{\atom{a}}_i}{d_i} \;\land\;
  I \subseteq J \;\land\; 
  \forall i \in I.\; c_i \compliant d_i
\\
\nonumber
  & \hspace{8pt} \lor
  && d  = \sumI{\bar{\atom{a}}_i}{d_i} \;\land\;
  I = \setenum{i} \;\land\; 
  c_i \compliant d_i
\\[10pt]
\nonumber
  & \hspace{8pt} \lor
  && d  = \ready{\atom{b}}.d' \;\land\;
  c \compliant d'
\\[10pt]
\label{eq:compliant-inversion:ext}
  c = \SumExt[i \in I]{\atom{a}_i}{c_i'} 
  \quad
  & \implies
  && d = \SumInt[i \in J]{\bar{\atom{a}}_i}{d_i} \;\land\;
  J \subseteq I \;\land\; 
  \forall i \in J.\; c_i \compliant d_i
\\
\nonumber
  & \hspace{8pt} \lor
  && d  = \SumExt[i \in J]{\bar{\atom{a}}_i}{d_i} \;\land\;
  I \cap J \neq \emptyset \;\land\; 
  \forall i \in I \cap J.\; c_i \compliant d_i
\\
\nonumber
  & \hspace{8pt} \lor
  && d  = \ready{\atom{b}}.d' \;\land\;
  c \compliant d'
\\[10pt]
\label{eq:compliant-inversion:rdy}
  c = \ready{\atom{a}}.c'
  & \implies
  && c' \compliant d
\end{align}
\end{subequations}
\end{applemma}
\begin{proof}
For~\eqref{eq:compliant-inversion:rdy}, 
assume that $c = \ready{\atom{a}}.c'$.
By rule \nrule{[Rdy]}, 
\(
{\pmv A} \says c \mid {\pmv B} \says d \cmove{{\pmv A} \says \atom{a}}
{\pmv A} \says c' \mid {\pmv B} \says d
\).
Since $c \compliant d$, by item (2) of Def.~\ref{def:compliance}
it must be the case that
$c' \compliant d$.

\smallskip\noindent
For~\eqref{eq:compliant-inversion:int}, 
let $c = \SumInt[i \in I]{\atom{a}_i}{c_i}$.
We have three subcases, according to the form of $d$.
\begin{itemize}

\item $d = \SumExt[i \in J]{\atom{b}_i}{d_i}$.
We start by proving that 
$\setenum{\atom{b}_i}_{i \in J} \supseteq \co{\setenum{\atom{a}_i}_{i \in I}}$.
Let $\atom{a} \in \setenum{\atom{a}_i}_{i \in I}$.
By Def.~\ref{def:rs}, we have that $\setenum{\atom{a}} \in \rs{c}$.
Since $c \compliant d$ and $c,d$ are $\ready{}\!$-free, 
by Def.~\ref{def:compliance} we have that
$\setenum{\atom{a}} \cap \co{\mathcal{Y}} \neq \emptyset$ for all $\mathcal{Y} \in \rs{d}$.
By Def.~\ref{def:rs}, $\rs{d} = \setenum{\atom{b}_i}_{i \in J}$.
Therefore, there exists $i \in J$ such that $\atom{b}_i = \atom{a}$.

We now prove that $c_i \compliant d_i$, for all $i \in I$.
Let $j \in I$. 
By rule [\nrule{IntExt}], we have that
\(
{\pmv A} \says c \mid {\pmv B} \says d \cmove{{\pmv A} \says \atom{a}_j}
{\pmv A} \says c_j \mid {\pmv B} \says \ready{\bar{\atom{a}}_j}.d_j
\).
By item (2) of Def.~\ref{def:compliance} it follows that
$c_j \compliant \ready{\bar{\atom{a}}_j}.d_j$.
Therefore, by~\eqref{eq:compliant-inversion:rdy} we conclude that
$c_j \compliant d_j$.

\item $d = \SumInt[i \in J]{\atom{b}_i}{d_i}$.
We start by proving that $|I| = |J| = 1$.
If this were not the case, then we could apply rule [\nrule{IntIntFail}],
and so by Lemma~\ref{lem:compliant-fail} we would have the contradiction
$c \ncompliant d$.
Therefore, let $I = J = \setenum{j}$.
By rule [\nrule{IntInt}] we have that
\(
{\pmv A} \says c \mid {\pmv B} \says d \cmove{{\pmv A} \says \atom{a}_j}
{\pmv A} \says c_j \mid {\pmv B} \says \ready{\bar{\atom{a}}_j}.d_j
\).
By item (2) of Def.~\ref{def:compliance} it follows that
$c_j \compliant \ready{\bar{\atom{a}}_j}.d_j$.
Therefore, by~\eqref{eq:compliant-inversion:rdy} we conclude that
$c_j \compliant d_j$.

\item $d = \ready{\atom{b}}.d'$. 
By~\eqref{eq:compliant-inversion:rdy} 
we conclude that $c \compliant d'$.

\end{itemize}

\smallskip\noindent
For~\eqref{eq:compliant-inversion:ext}, 
let $c = \SumExt[i \in I]{\atom{a}_i}{c_i}$.
We have three subcases, according to the form of $d$.
\begin{itemize}

\item $d = \SumInt[i \in J]{\atom{b}_i}{d_i}$.
This case has been already dealt with 
when proving~\eqref{eq:compliant-inversion:int}
in the case where $d$ is an external sum.

\item $d = \SumExt[i \in J]{\atom{b}_i}{d_i}$.
We start by proving that
$\setenum{\atom{b}_i}_{i \in J} \cap \co{\setenum{\atom{a}_i}_{i \in I}} \neq \emptyset$.
If this were not the case, then we could apply rule [\nrule{ExtExtFail}],
and so by Lemma~\ref{lem:compliant-fail} we would have the contradiction
$c \ncompliant d$.
We now prove that $c_i \compliant d_i$, for all $i \in I \cap J$.
Let $j \in I \cap J$. 
By rule [\nrule{ExtExt}], we have that
\(
{\pmv A} \says c \mid {\pmv B} \says d \cmove{{\pmv A} \says \atom{a}_j}
{\pmv A} \says c_j \mid {\pmv B} \says \ready{\bar{\atom{a}}_j}.d_j
\).
By item (2) of Def.~\ref{def:compliance} it follows that
$c_j \compliant \ready{\bar{\atom{a}}_j}.d_j$.
Therefore, by~\eqref{eq:compliant-inversion:rdy} we conclude that
$c_j \compliant d_j$.
 
\item $d = \ready{\atom{b}}.d'$. 
By~\eqref{eq:compliant-inversion:rdy} 
we conclude that $c \compliant d'$.
\qed
\end{itemize}
\end{proof}

\begin{applemma} \label{lem:culpable-inversion}
For all $\gamma$, $\cfrown{\pmv A}{\gamma}$, if and only if $\gamma$ has one of 
the following forms:
\begin{align*}
& \bic{\cnil}{d} 
&&
\tag*{[\nrule{\frownie Nil}]} \\[7pt]
& \bic{\ready{\atom{a}} .\; c}{d} 
&& \text{ with } \atom{a} \neq \atom{e}
\tag*{[\nrule{\frownie Rdy}]} \\[7pt]
& \bic{\SumInt[i \in I]{\atom{a}_i}{c_i}}{d}
&& \text{ with } \forall i \in I.\; \atom{a}_i \neq \atom{e} \text{ and } d \;\ready{}\!\text{-free}
\tag*{[\nrule{\frownie Int}]} \\
& \bic{\SumExt[i \in I]{\atom{a}_i}{c_i}}{\SumExt[j \in J]{\atom{b}_j}{d_j}}
&& \text{ with } \begin{array}{l}
  \atom{e} \not\in \setenum{\atom{a}_i}_{i \in I}, \;\mathit{or}\; \\
  \atom{e} \not\in (\setenum{\atom{a}_i}_{i \in I} \cap \co{\setenum{\atom{b}_j}_{j \in J}}) \neq \emptyset
  \end{array}
\tag*{[\nrule{\frownie Ext}]}
\end{align*}
\end{applemma}
\begin{proof}
For $(\Leftarrow)$, the proof is by straightforward case analysis,
using the rules for $\cmove{}$.
\begin{itemize}

\item if $\gamma$ has the form \nrule{[\frownie Nil]}, then
$\cfrown{\pmv A}{\gamma}$ follows directly by Def.~\ref{def:culpable};

\item if $\gamma$ has the form \nrule{[\frownie Rdy]}, then
by the rule \nrule{[Rdy]} $\gamma$ can take a transition 
labelled ${\pmv A} \says \atom{a}$, with $\atom{a} \neq \atom{e}$;
no other transitions are possible.

\item if $\gamma$ has the form \nrule{[\frownie Int]}, 
assume that $I \neq \emptyset$ (case already dealt with).
Then, one of the rules 
\nrule{[IntExt]}, \nrule{[IntExtFail]},
\nrule{[IntInt]}, \nrule{[IntIntFail]}, 
allow $\gamma$ to take a transition 
labelled ${\pmv A} \says \atom{a}_i$, with $\atom{a}_i \neq \atom{e}$.

\item if $\gamma$ has the form \nrule{[\frownie Ext]}, 
assume that $I \neq \emptyset$ (case already dealt with).
There are two subcases.
If $\atom{e} \not \in \setenum{\atom{a}_i}_{i \in I}$, then
one of the rules \nrule{[ExtExt]} or \nrule{[ExtExtFail]}
allow $\gamma$ to take a transition labelled ${\pmv A} \says \atom{a}_i$,
but no transition labelled ${\pmv A} \says \atom{e}$.
If  $\atom{e} \in \setenum{\atom{a}_i}_{i \in I}$, 
$\atom{e} \not \in \setenum{\atom{b}_j}_{j \in J}$, 
and $(\setenum{\atom{a}_i}_{i \in I} \cap \co{\setenum{\atom{b}_j}_{j \in J}}) \neq \emptyset$,
then no transitions labelled ${\pmv A} \says \atom{e}$ are possible, but
there exists a transition labelled ${\pmv A} \says \atom{a}_i$.

\end{itemize}

For $(\Rightarrow)$, assume that $\gamma$ has none of the forms
reported in the statement.
We will prove that $\csmiley{\pmv A}{\gamma}$.
We proceed by cases on the form of $\gamma$.
\begin{itemize}

\item $\bic{\ready{\atom{e}}.c'}{d}$.
By rule \nrule{[Rdy]}, there is a transition labelled ${\pmv A} \says \atom{e}$.

\item $\bic{c}{d}$, with 
$c = \SumInt[i \in I]{\atom{a}_i}{c_i}$, and
$d = \ready{\atom{b}}.d'$ or $\exists i \in I.\; \atom{a}_i = \atom{e}$.
\begin{itemize}

\item If $d = \ready{\atom{b}}.d'$, then the only possible transition, 
obtained by the rule \nrule{[Rdy]}, 
is labelled ${\pmv B} \says \atom{b}$.

\item If $\exists i \in I.\; \atom{a}_i = \atom{e}$, then
a transition labelled ${\pmv A} \says \atom{e}$ is possible by using one
of the rules
\nrule{[IntExt]}, \nrule{[IntExtFail]},
\nrule{[IntInt]}, \nrule{[IntIntFail]}.

\end{itemize}

\item $\bic{c}{d}$, 
with $c = \SumExt[i \in I]{\atom{a}_i}{c_i}$
we have the following subcases:
\begin{itemize}

\item if $d = \ready{\atom{b}}.d'$ or $d = \SumInt[j \in J]{\atom{b}_j}{d_j}$, 
with $J \neq \emptyset$, then $\gamma$ cannot take {\pmv A}-transitions.

\item if $d = \SumExt{\atom{b}_i}{d_i}$ and
$\atom{e} \in \setenum{\atom{a}_i}_{i \in I}$,
we have two subcases.
If $\atom{e} \in \setenum{\atom{b}_j}_{j \in J}$,
then \nrule{[ExtExt]} allows for a transition
labelled ${\pmv A} \says \atom{e}$.
If $(\setenum{\atom{a}_i}_{i \in I} \cap \co{\setenum{\atom{b}_j}_{j \in J}}) = \emptyset$, a transition with the same label is obtained by \nrule{[ExtExtFail]}.
\qed
\end{itemize} 

\end{itemize}
\end{proof}

\begin{barttheorem}{Theorem~\ref{lem:compliant-smiley}}
\lemcompliantsmiley
\end{barttheorem}
\begin{proof}
We first consider the case that $c$ is $\ready{}\!$-free,
where we have the following three exhaustive subcases:
\begin{itemize}

\item $\gamma \cmove{{\pmv A} \says \atom{a}} \delta$,
for some $\atom{a}$, 
and the transition has been possible through one of the rules
\nrule{[IntExt]}, \nrule{[IntInt]}, or \nrule{[ExtExt]}
in Figure~\ref{fig:contracts:semantics:nonfail}.
Then, the contract advertised by {\pmv B} in 
$\delta$ will have the form $\ready{\bar{\atom{a}}}.d'$, 
for some $d'$, while the contract of {\pmv A} is $\ready{}\!$-free.
By Lemma~\ref{lem:culpable-inversion}
we have that $\csmiley{\pmv A}{\delta}$ (since $c$ is $\cnil$-free).
Therefore, the thesis follows by choosing $\gamma' = \delta$ 
and $\eta = ({\pmv A} \says \atom{a})$.

\item $\gamma \cmove{{\pmv A} \says \atom{a}} \delta$,
for some $\atom{a}$, 
and the transition has been possible through one of the rules
\nrule{[--Fail]} in Figure~\ref{fig:contracts:semantics:fail}.
Then, $\delta = \bic{\E}{\cnil}$, and 
by Definition~\ref{def:culpable}
we have that $\csmiley{\pmv A}{\delta}$.
The thesis follows by choosing $\gamma' = \delta$
and $\eta = ({\pmv A} \says \atom{a})$.

\item $\gamma$ cannot take a transition under an action of {\pmv A}.

By Definition~\ref{def:culpable}
we have that $\csmiley{\pmv A}{\gamma}$. 
Therefore, the thesis follows with
$\gamma' = \gamma$ (and empty $\eta$).

\end{itemize}

We now consider the case $c = \ready{\atom{a}}.c'$, 
for some $\atom{a}$ and $c'$.
By rule \nrule{[Rdy]}
$\gamma$ has a transition to $\bic{c'}{d}$
labelled ${\pmv A} \says \atom{a}$.
Since $c'$ is $\ready{}\!$-free and $\cnil$-free, 
we then apply one of the three cases above 
(which guarantee $|\eta| \leq 1$)
and conclude.
\qed
\end{proof}

\begin{applemma} \label{lem:double-smiley}
Let 
\(
  \gamma_0 \cmove{{\pmv A} \says \atom{a}} 
  \gamma_1 \cmove{{\pmv A} \says \atom{b}} 
  \gamma_2 = \bic{c_2}{d_2}
\). 
Then,
$c_2 = \cnil$ or $\csmiley{\pmv A}{\gamma_2}$.
\end{applemma}
\begin{proof}
We first prove that, for all ${\gamma}$:
\begin{equation}  \label{eq:double-smiley}
  {\gamma} \;\;\ready{}\!\textit{-free}
  \;\land\;
  {\gamma} \cmove{{\pmv A} \says -} 
  {\gamma}' = \bic{{c}'}{{d}'}
  \implies
  {c}' = \cnil 
  \;\lor\; 
  \csmiley{\pmv A}{{\gamma}'}
\end{equation}
Let ${\gamma} = \bic{{c}}{{d}}$.
We have the following two subcases:
\begin{itemize}

\item ${\gamma} \cmove{{\pmv A} \says -} {\gamma}'$
has been derived through one of the rules
in Figure~\ref{fig:contracts:semantics:nonfail}
(except \nrule{[Rdy]}).
Then, $d'$ will have the form $\ready{\bar{\atom{a}}}.\tilde{d}$, 
for some $\tilde{d}$, while $c'$ is $\ready{}\!$-free.
Then, either ${c}' = \cnil$, or
by Lemma~\ref{lem:culpable-inversion}
we have that $\csmiley{\pmv A}{{\gamma}'}$.

\item ${\gamma} \cmove{{\pmv A} \says -} {\gamma}'$ 
has been derived through one of the rules
in Figure~\ref{fig:contracts:semantics:fail}.
Then, ${\gamma}' = \bic{\E}{\cnil}$, and 
by Definition~\ref{def:culpable}
we have that $\csmiley{\pmv A}{{\gamma}'}$.

\end{itemize}

Back to the main statement, let $\gamma_i = \bic{c_i}{d_i}$,
for $i \in \setenum{0,1,2}$.
After the first transition, we have that
$c_1 \neq \cnil$, because otherwise the transition to $\gamma_2$
would not be possible.
Also, $c_1$ is $\ready{}\!$-free, because a transition labelled
${\pmv A} \says \atom{a}$ cannot generate a $\ready{}$ in {\pmv A}.
Therefore, by the hypothesis $\gamma_1 \cmove{{\pmv A} \says \atom{b}} \gamma_2$
and by~\eqref{eq:double-smiley}
it follows that either $c_2 = \cnil$ or $\csmiley{\pmv A}{\gamma_2}$.
\qed
\end{proof}

\begin{barttheorem}{Theorem~\ref{lem:compliant-frown}}
\lemcompliantfrown
\end{barttheorem}
\begin{proof}
Assume that $\csmiley{\pmv A}{\gamma}$ and $\csmiley{\pmv B}{\gamma}$.
According to Lemma~\ref{lem:culpable-inversion}, 
$\gamma$ must have one of the following forms
(symmetric cases are omitted):
\begin{itemize}

\item ${\pmv A} \says c' \mid {\pmv B} \says \ready{\atom{e}}.E$. 
In this case, $\gamma$ has been obtained through a transition step labelled
${\pmv A} \says \atom{e}$.
By Definition~\ref{def:contracts:semantics} and by the syntactic restriction
on the continuations of $\atom{e}$, 
this implies $c' = E$.
By Definition~\ref{def:contracts:syntax},
$c' = E \equiv \sumI{\atom{e}}{X}\setenum{\bind{X}{E}} \equiv \sumI{\atom{e}}{E}$
succeeds, as well as $d' = \ready{\atom{e}}.E$.

\item $c' = \sumI{\atom{e}}{E} \sumInt c_1$ and 
$d' = \sumI{\atom{e}}{E} \sumInt d_1$.
By Definition~\ref{def:contracts:syntax}, both $c'$ and $d'$ succeed.

\item $c' = \sumI{\atom{e}}{E} \sumInt c_1$ and 
$d' = \SumExt[i \in I]{\atom{b}_i}{d_i}$.
Since $c' \compliant d'$ and $\setenum{\atom{e}} \in \rs{c'}$, 
then there exists $j \in I$ such that
$\atom{b}_j = \bar{\atom{e}} = \atom{e}$.
By Definition~\ref{def:contracts:syntax}, both $c'$ and $d'$ succeed.

\item $c' = \SumExt[i]{\atom{a}_i}{c_i}$ and 
$d' = \SumExt[j]{\atom{b}_i}{d_j}$,
with $\atom{e} \in \setenum{\atom{a}_i}_i \cap \setenum{\atom{b}_j}_j$.
By Definition~\ref{def:contracts:syntax}, both $c'$ and $d'$ succeed.

\item $c' = \SumExt[i]{\atom{a}_i}{c_i}$ and 
$d' = \SumExt[j]{\atom{b}_i}{d_j}$,
with $\atom{e} \not\in \setenum{\atom{a}_i}_i \cup \setenum{\atom{b}_j}_j$,
and $\setenum{\atom{a}_i}_i \cap \co{\setenum{\atom{b}_j}_j} = \emptyset$.
The latter condition violates requirement $(1)$ 
of Definition~\ref{def:compliance}, so it is false that $c' \compliant d'$ 
--- contradiction.
\qed

\end{itemize}
\end{proof}

\section{Proofs for Section~\ref{sect:honesty}}

\begin{barttheorem}{Lemma~\ref{lem:readydo}}
\lemreadydo
\end{barttheorem}
\begin{proof}
Let $\mathcal{X} = \readydo{s}{P}$, and
assume that $\cfrown{\pmv A}{\gamma}$.
We have the following cases on the structure of $c$:
\begin{itemize}

\item $c = \cnil$. 
This case never applies, because by Def.~\ref{def:unblocks},
$\unblocks{\cnil}{\mathcal{Y}}$ is false for all $\mathcal{Y}$.

\item $c = \ready{\atom{a}}.c'$. 
Since $\cfrown{\pmv A}{\gamma}$, 
by Def.~\ref{def:culpable} it must be $\atom{a} \neq \atom{e}$.
Since $\unblocks{c}{\mathcal{X}}$, then by Def.~\ref{def:unblocks}
$\atom{a} \in \mathcal{X}$.
So, by Def.~\ref{def:readydo} there exists an unguarded 
$\fact{s}{\atom{a}}$ in $P$.
Since $\gamma \cmove{{\pmv A} \says \atom{a}}$ by \nrule{[Rdy]},
then by the rules \nrule{[Do]} and \nrule{[Par]},
$S \pmove{{\pmv A} \says \fact{s}{\atom{a}}}$.

\item $c = \SumInt[i \in I]{\atom{a}_i}{c_i}$. 
Since $\cfrown{\pmv A}{\gamma}$, 
by Lemma~\ref{lem:culpable-inversion} it must be $\atom{a}_i \neq \atom{e}$
for all $i \in I$.
Since $\unblocks{c}{\mathcal{X}}$, then by Def.~\ref{def:unblocks}
there exists $i \in I$ such that $\atom{a}_i \in \mathcal{X}$.
So, by Def.~\ref{def:readydo} there exists an unguarded 
$\fact{s}{\atom{a}_i}$ in $P$.
Note that $d$ is $\ready{\!}$-free, since otherwise it would be
$\csmiley{\pmv A}{\gamma}$.
Therefore, $\gamma \cmove{{\pmv A} \says \atom{a}_i}$
by either \nrule{[IntExt]} or \nrule{[IntExtFail]}, and then
$S \pmove{{\pmv A} \says \fact{s}{\atom{a}_i}}$
by \nrule{[Do]} and \nrule{[Par]}.

\item $c = \SumExt[i \in I]{\atom{a}_i}{c_i}$. 
Since $\cfrown{\pmv A}{\gamma}$ and $I \neq \emptyset$, 
choose $j \in I$ such that $\gamma \cmove{{\pmv A} \says \atom{a}_j}$.
Since $\unblocks{c}{\mathcal{X}}$ and $\atom{a}_j \neq \atom{e}$, 
then by Def.~\ref{def:unblocks}, $\atom{a}_j \in \mathcal{X}$.
So, by Def.~\ref{def:readydo} there exists an unguarded 
$\fact{s}{\atom{a}_j}$ in $P$.
Then by the rules \nrule{[Do]} and \nrule{[Par]},
$S \pmove{{\pmv A} \says \fact{s}{\atom{a}_j}}$.
\qed

\end{itemize}

\end{proof}

\begin{barttheorem}[ (Factual exculpation)]{Theorem~\ref{lem:doubledo}}
\lemdoubledo
\end{barttheorem}
\begin{proof}
Straightforward by Lemma~\ref{lem:double-smiley},
by noting that the steps from $\mu_i$ to $\mu_{j-2}$
do not change the contract in $s$, while the steps
$\mu_{i-1}$ and $\mu_{j-1}$ correspond to $\cmove{}$-transitions 
labelled ${\pmv A} \says \atom{a}$ and ${\pmv A} \says \atom{b}$, 
respectively.
\qed
\end{proof}

We now prove that honesty is undecidable.
To do that, we show that the complement problem, i.e.\ deciding if
a participant is \emph{dishonest}, is not recursive 
(actually, it is recursively enumerable, from which it follows that
honesty is neither recursive nor recursively enumerable). 

\begin{barttheorem}{Theorem~\ref{th:honesty-undecidable}}
\thhonestyundecidable
\end{barttheorem}
\begin{proof}
We start by proving that ``$\sys {\pmv A} P$ dishonest'' is a r.e.\ property. 
By Def.~\ref{def:honest}, $\sys {\pmv A} P$ is \emph{not} honest iff 
there exists a context $S$ (free from latent/stipulated contracts of {\pmv A}) 
such that {\pmv A} is not honest in $\sys {\pmv A} P \mid S$. 
The latter holds when there exist a contract $c$ and a session $s$ such that 
{\pmv A} does not realize $c$ at $s$ in $\sys {\pmv A} P \mid S$. 
Summing up, $\sys {\pmv A} P$ is dishonest iff the following conditions
hold for some $S,S_0,s$:
\begin{enumerate}
\item $S$ free from ${\pmv A} \says \cdots$ and from $\sys {\pmv A} \cdots$
\item $\sys {\pmv A} P \mid S \pmove{}^* S_0$
\item there is an {\pmv A}-solo $\setenum{{\pmv A} \says \pi}$-fair 
trace $S_0 \pmove{} S_1 \pmove{} \cdots$ where 
$\cfrown[s]{\pmv A}{S_j}$ for all $j \geq 0$.
\end{enumerate}

Recall that $p(x,y)$ r.e.\ implies that $q(y) = \exists x. p(x,y)$ 
is \text{r.e.}, provided that $x$ ranges over an effectively enumerable set
(\text{e.g.}, systems $S$, or sessions $s$).  
Thus, to prove the above existentially-quantified property r.e.\ 
it suffices to prove that 1), 2), 3) are \text{r.e.}.  
Property 1 is trivially recursive.  
Property 2 is r.e.\ since one can enumerate all the possible finite traces.  
Property 3 is shown below to be recursive, by reducing it to the
satisfiability of a LTL formula on Petri Nets.

Deciding property 3 amounts to deciding the satisfiability
of the LTL property $\Box (\cfrown{\pmv A}{})$ on the 
(fair) LTS generated from $S_0$. 
Note that $\cfrown[]{\pmv A}{}$ is a decidable property, 
and we need to consider the {\pmv A}-solo fair traces, only.
The fairness requirement can be moved from the LTS into the
formula itself: indeed, the satisfiability of $\Box (\cfrown{\pmv A}{})$ 
on the fair traces is equivalent to the satisfiability of
$\mathit{fair} \land \Box (\cfrown[s]{\pmv A}{})$ on all the traces,
where $\it fair$ encodes the fairness requirement in LTL.
In order to check the latter, first note that restricting
to {\pmv A}-solo traces allows us to neglect all the interactions 
with the context. 
Also, participant {\pmv A} can only interact with a finite, statically
bounded number of sessions: she needs to consume a latent contract
from another participant to spawn a fresh session, and those must
be present in $S_0$. 
Because of this, without loss of generality, it is possible to 
assume that the continuation 
of each \coco prefix is a defined process $X_i(\vec{u})$ 
where $i$ ranges over a finite, statically known set. 
Further, $\vec{u}$ can only be instantiated in a finite number of ways, 
since there are only so many sessions.  
This makes the process $Q$ in $\sys {\pmv A} Q$ 
equivalent to a parallel composition of such $X_i(\cdots)$, 
each one possibly occurring zero, one or more times. 
Therefore, the systems $\sys {\pmv A} Q$ can be equivalently represented as
a Petri net, where places are $X_i$ and tokens account for their
multiplicity ($\tell{}{}$ actions of {\pmv A} to advertise contracts to 
the context are immaterial,
since they cannot be fused in an {\pmv A}-solo trace). 
The outer context of $\sys {\pmv A} Q$ in system $S_0$ is 
a finite-state system. 
Indeed, sessions appear in a statically bounded number, 
and each one of them has a finite-state contract; 
further, participants other than A can not move
In an {\pmv A}-solo trace).

To conclude, property 3 reduces to the problem of 
model checking LTL over Petri nets, 
which is decidable~\cite{Esparza94decidability}.


We now prove that the property ``$\sys {\pmv A} P$ is dishonest'' is
undecidable.
To do that, we reduce the halting problem on Turing machines to the
problem of checking dishonesty.
We model a configuration of a generic Turing machine $M$ as a finite sequence
\[ x_0\ x_1\ x_2\ \ldots (x_n , q) \ldots  x_k \]
where 
\begin{enumerate}
\item \label{1} $x_i$ represents the symbol written at the $i$-th cell
  of the tape,
\item \label{2} the single occurrence of the pair $(x_n, q)$ denotes
  that the head of $M$ is over cell $n$, and $M$ is in state $q$,
\item \label{3} the tape implicitly contains ``blank'' symbols at
  cells after position $k$.
\end{enumerate}
Without loss of generality, assume that $M$ halts only when its head
is over $x_0$ and $M$ is in a halting state $q_\mathrm{stop}$.
Note that each symbol $x_i$ can be finitely represented, because the
alphabet is finite. 
States $q$ can be finitely represented as well.

Given a deterministic, 1-tape Turing Machine $M$, we devise an
effective procedure to construct a participant which is dishonest if and only
if $M$ halts on the empty tape.
The system has the form
\begin{equation}\label{eq:AM}
\sys {\pmv A}{(x) \tell B {\freeze x c} .  \fact x {\atom a} . P}
\end{equation}
where the choice of names for participant $\pmv B$ and atom $\atom a$
is immaterial, $c = \rec X {{\atom a};X}$, and $P$ is given below.
Note that we will \emph{not} construct a participant $\pmv A$ which
simulates $M$ by herself; rather, in order to simulate $M$, $\pmv A$
will require some cooperation from the context.
So, we guarantee two different properties according to whether the context
cooperates:
\begin{itemize}
\item if $\pmv B$ does not cooperate, $\pmv A$ will stop simulating
  $M$, but will still behave honestly in all the open sessions;
\item if $\pmv B$ cooperates, $\pmv A$ will run $M$, and behave
  honestly while doing that; only when $M$ is found to halt, $ \pmv A$
  will instead behave dishonestly.
\end{itemize}
In other words, if $M$ does not halt, $\pmv A$ is honest in all
contexts (and therefore honest); if $M$ halts, $\pmv A$ is not honest
in at least one (cooperating) context (and therefore dishonest).

We now define the process $P$ in the system~\eqref{eq:AM} (hereafter,
we denote with $s$ the session name instantiated for $x$).
Such process is defined so that whenever $M$ halts, ${\pmv A}$ will be
dishonest at $s$ and ${\pmv A}$ will be otherwise honest at all her
sessions (including $s$).
Note that if the latent contract $\freeze x c$ in~\eqref{eq:AM} is
never fused by $\pmv B$, then $\pmv A$ is honest.

By the finiteness conditions on $M$, we can represent the information
relative to a single cell through a finite contract family
$d_{x,q}$ in which $x$
ranges over the alphabet, and $q$ over the states (plus one extra
element, representing the fact that the head is elsewhere).
More precisely, $d_{x,q}$ is defined as
\[
d_{x,q} = \atom{read}_{x,q} ; d_{x,q} 
\sumInt \SumInt[x']{\atom{writeSymbol}_{x'}}{d_{x',q}}
\sumInt \SumInt[{q'}]{\atom{writeState}_q'}{d_{x,q'}}
\]
where $\atom{read}_{x,q}$, $\atom{writeSymbol}_x$, $\atom{writeState}_q$ are atoms.
%
Note in passing that mutual recursion can be reduced to single
recursion via the $\mathit{rec}$ construct (up to some unfolding,
Beki\'c's Theorem).

Process $P$ in~\eqref{eq:AM} uses the above contracts in separate
sessions, one for each tape cell.
Informally, $P$ is built so to generate
\[
\mathit{Begin}(s_0,s_1) \mid X(s_0,s_1,s_2) \mid X(s_1,s_2,s_3) \mid \ldots \mid \mathit{End}(s_{n-1},s_n)
\]
where $s_0,\ldots,s_n$ are distinct sessions.
Processes $\mathit{Begin}$, $X$, and $\mathit{End}$ are constructed so
to behave as follows:
\begin{itemize}
\item $\mathit{Begin}(s_0,s_1)$ handles the leftmost cell of the tape.
  It behaves as $X(\_)$ (defined below), but also keeps on performing
  $\fact s a$, hence realizing the first stipulated contract $c$.
  Process $\mathit{Begin}(s_0,s_1)$ also waits for the head of $M$ to
  reach the leftmost cell and $M$ to be in the halting state
  $q_{\mathrm{stop}}$.
  When this is detected, it stops performing $\fact s a$, hence making
  $\pmv A$ become dishonest (at session $s$).
\item Processes $X(\_,s_i,\_)$ are responsible for handling the $i$-th
  cell.
  Each such process reads the cell by performing $\sum_{x,q}
  \fact{s_i}{\atom{read}_{x,q}} . \mathit{Handle}_{x,q}$.
  Whenever the head is \emph{not} on the $i$-th cell, it keeps on
  performing reads (so that $\pmv A$ does not become culpable at
  session $s_i$).
  If the head is on the $i$-th, the symbol is updated according to the
  transition rules, and then the head is possibly moved.
  Moving the head requires performing a
  $\fact{s_j}{\atom{writeState}}$ where $j$ is either $i-1$ or $i+1$.
\item Process $\mathit{End}(s_{n-1},s_n)$ waits for the head to reach
  the $n$-th cell.
  When this happens, it creates a new session $s_{n+1}$ (by issuing a
  $\tell B {}$ of a frozen, which may be possibly fused by B
  --otherwise $\pmv A$ remains honest in all sessions), spawns a new
  process $X(s_{n-1},s_n,s_{n+1})$, and then recurse as
  $\mathit{End}(s_n,s_{n+1})$.
\end{itemize}
A crucial property is that it is possible to craft the above processes
so that in no circumstances (including hostile contexts)
they make $\pmv A$ dishonest at sessions $s_i$; the only
session where $\pmv A$ could eventually become culpable is $s$.
For example, $X(\_,s_i,\_)$ is built so it never stops performing
reads at session $s_i$. 
This property is achieved by encoding each potentially blocking
operation $\fact{s_k} b .\;  P'$ as $Q = \fact{s_k} b . P' +
  \sum_{x,q}
  \fact{s_i}{\atom{read}_{x,q}} . Q$.
Indeed, in this way, reads on $s_i$ are continuously
performed, unless the context suddenly stops cooperating in that session:
in this case, the context is culpable in $s_i$, but
$\pmv A$ is not. Also, in this case the computation of $M$ may get stuck,
but $\pmv A$ would still be honest in every session, as intended.
$\pmv A$ could also get stuck she is waiting to write at session $s_{i+1}$.
While performing the write action, care must be taken so to
not forget to keep on reading on $s_i$, preserving honesty at $s_i$.
When that is done, even if the other participant involved at 
session $s_{i+1}$ is making such session stuck, $\pmv A$ 
keeps moving at $s_i$.
Similarly, the $\mathit{End}$ process is built so to keep on 
reading from $s_n$ when waiting for a new session $s_{n+1}$ to be opened.
In the case the context does not provide a compliant latent contract
and the session can not be spawned, this may stop the computation of $M$,
but $A$ will still be honest in all the opened sessions $s_i$.

To conclude, given a Turing Machine $M$ we have constructed a \coco
participant $\sys {\pmv A} P$ such that $(i)$ if $M$ does not halt,
then $\sys {\pmv A} P$ is honest, while $(ii)$ if $M$ halts, then $\sys
{\pmv A} P$ is not honest in some (cooperating) context.
Note that a context which cooperates with $\sys {\pmv A} P$ always
exists: e.g., the system that first tells the duals of all the
contracts possibly advertised by ${\pmv A}$ (a finite number), fuses
them, and then (recursively) performs all the promised actions.
\qed
\end{proof}


\section{Proofs for Section~\ref{sect:realizability}}

\begin{barttheorem}{Lemma~\ref{lem:abstract-contract}}
  \lemabstractcontract
\end{barttheorem}
\begin{proof}
  The first item is straightforward by case analysis on the rules
  in Def.~\ref{def:contracts:semantics} and Def.~\ref{def:abstract-contract}.
  The second item is similar, by also exploiting the fact that, 
  since $c \compliant d$, 
  Lemma~\ref{lem:compliant-fail} prevents from $\abscmove{\cnil}$ transitions.
\qed
\end{proof}

\begin{applemma} \label{lem:ctx-move}
  $ c \abscmove{ctx} c' \abscmove{\atom{a}} c''
  \implies
  c \abscmove{\atom{a}} c''$
\end{applemma}
\begin{proof}
  By inspection of the rules for $\abscmove{}$. Indeed when $c'$
  differs from $c$, this is due to having selected a specific
  branch in an external sum, or to having a $\ready{\atom{a}}$ prefix
  instead of a singleton internal sum. In all cases $\abscmove{\atom{a}}$
  leads to the same result.
\qed
\end{proof}

\begin{applemma} \label{lem:realizes-ctx}
  For all processes $Q$ and for all contracts $c$:
  \[
  \realizes[\pmv A]{Q}{c}{s}
  \;\land\;
  c \abscmove{ctx} c'
  \;\;\implies\;\;
  \realizes[\pmv A]{Q}{c'}{s}
  \]
\end{applemma}
\begin{proof}
  We define a relation $\mathcal{R}$ as follows
  \[
  P \mathcal{R} c \iff
  \realizes{P}{c}{} \lor 
     (\exists c'.\ \realizes{P}{c'}{} \land c' \abscmove{ctx} c)
  \]
  and then prove that it satisfies the conditions of
  Definition~\ref{def:realizes}. This would imply that $\mathcal{R}$ concides
  with $\sharp$-realizability, hence the statement of the lemma holds.

  When $P \mathcal{R} c$ is due to $\sharp$-realizability, clearly it satisfies
  the required conditions. The non trivial case is when
  for some $c'$ we have $\realizes{P}{c'}{}$ and $c' \abscmove{ctx} c$.
  To check the conditions in this case, let
  $P=P_0,\ldots$ be a $\setenum{\unblocked,\tell{}}$-fair do-free trace.

  We proceed by cases on the transition $c' \abscmove{ctx} c$:

  \begin{itemize}
  \item We have $c' \abscmove{ctx} c=c'$. Here, we get the conditions from
    $\realizes{P}{c}{}=c'$.

  \item We have $c' = \SumExt{\atom{a}_i}{c_i} \abscmove{ctx} 
    \ready{\atom{a}_n} \;.\; c_n = c$.
    Here, $c' \ unblocks\ \readydo{s}{P_j}$ as long as $j$ is sufficently large.
    This implies that for such $j$, forall $i$ we have
    $\atom{a}_i \in \readydo{s}{P_j}\cup\setenum{\atom{e}}$.
    Hence, $\atom{a}_n \in \readydo{s}{P_j}\cup\setenum{\atom{e}}$,
    which proves $c = \ready{\atom{a}_n} \;.\; c_n \ unblocks\ \readydo{s}{P_j}$.

    For the second condition, assume $P_j \abspmove{\fact{s}{\atom a}} P'$
    and that $c \abscmove{\atom{a}} c''$. Since $c' \abscmove{ctx} c$,
    by Lemma~\ref{lem:ctx-move} we get 
    $c' \abscmove{\atom{a}} c''$. Since $\realizes{P}{c'}{}$,
    we have that $\realizes{P'}{c''}{}$, hence $P' \mathcal{R} c''$.

  \item We have 
    $c' = \sumI{a}{c''} \abscmove{ctx} \ready{\atom{a}} \;.\; c'' = c$.
    Here, we proceed similarly to the above case.
    We get $c' \ unblocks\ \readydo{s}{P_j}$ as long as $j$ is sufficently large.
    This implies that for such $j$, we have
    $\atom{a} \in \readydo{s}{P_j}\cup\setenum{\atom{e}}$.
    This proves $c = \ready{\atom{a}} \;.\; c'' \ unblocks\ \readydo{s}{P_j}$.

    The second condition follows exactly as per the previous case.
\qed
  \end{itemize}

\end{proof}

\begin{barttheorem}{Lemma~\ref{lem:co2-to-abstract}}
  \lemcocotoabstract
\end{barttheorem}
\begin{proof}

  We proceed by induction on the number of steps. The base case (empty
  trace) is trivial.  Otherwise, from the inductive hypothesis we
  obtain $Q_1 \abspmove{\mu_1} Q_2 \cdots$ where the $\mu_{i+1}$ are
  as in the statement above. We now conclude by proving $Q_0
  \abspmove{\mu_0} Q_1$, and its related property about $\mu_0$, by
  examining the possible cases for $S_0 \pmove{} S_1$. Note that in
  the stable trace the delimitations are brought to the top-level of
  $S_1$, i.e. in $(\vec{u}_1)$: this is done by the $\open{-}$
  operator in the definition of $\abspmove{}$, which we can therefore
  neglect below.

  \begin{itemize}
  \item $\pmv A$ did not move, but its context did.

    If some other participant performed a $\tell{\pmv A}{\freeze{x}{c}}$,
    then we have $Q_1 = \freeze x c \mid Q_0$. Then, by the abstract
    semantics rules we get $Q_0 \abspmove{ctx} Q_1$.

    Otherwise, if some other participant performed a $\fuse{x}{-}$,
    this can instantiate variable $x$ in the whole system to the fresh name $s$.
    In this case $Q_1 = Q_0 \setenum{\bind{x}{s}}$, and indeed
    by the abstract semantics rules we have
    $Q_0 \abspmove{ctx} Q_0 \setenum{\bind{x}{s}} = Q_1$.

    In the other cases, we have $Q_0=Q_1$, and $Q_0 \abspmove{ctx} Q_0
    \sigma = Q_1$ is obtained by taking $\sigma=id$.

  \item $\pmv A$ moved, firing prefix $\pi$.
    We consider two further subcases.

    \begin{itemize}

    \item $\pi=\tell{\pmv A}{\freeze x c}$. 
      This is possible when $Q_0 = \tell{\pmv A}{\freeze x c}.P + Q \mid R$.
      In this case the residual $Q_1$ is 
      $\freeze x {\pmv A \says c} \mid P \mid R$, and
      $Q_0 \abspmove{\pi} Q_1$ directly follows from the abstract semantics rules.

    \item $\pi\neq\tell{\pmv A}{-}$. This is possible
      when $Q_0 = \pi.P + Q \mid R$. In this case the residual $Q_1$
      must be of the form $(P \mid R)\sigma$, where $\sigma=id$ except
      when $\pi = \fuse{x}{\phi}$.
      In this case, $\sigma$ accounts for the resulting 
      variable instantiations. Finally, $Q_0 \abspmove{\pi} Q_1$
      follows from the abstract semantics rules.


    \end{itemize}
  \end{itemize}

  We now verify that if the concrete trace is fair,
  then the abstract trace is $\setenum{\unblocked,\tell{}{}}$-fair.
  Indeed, if $\abspmove{\unblocked}$ is enabled from a certain step onwards,
  say from $Q_k$, this means that there is an unguarded $\tau$ prefix in
  $Q_i$ for all $i\geq k$. In that case we would have that 
  $S_i \pmove{\pmv A \says \tau}$ for all $i\geq k$.
  Therefore, in the concrete trace eventually $\pmv A$ performs a $\tau$.
  Hence, in the abstract trace a $\unblocked$ is eventually performed.
  A similar reasoning applies to $\tell{}{}$.
  \qed
\end{proof}

\begin{applemma}\label{lem:realizes-step}
  If $\realizes[\pmv A]{Q}{c}{s}$, and $Q \abspmove[\pmv A]{\mu} Q'$, then

  \begin{enumerate}
  \item $\mu \neq \fact{s}{-} \implies \realizes[\pmv A]{Q'}{c}{s}$
    \label{eq:realizes-step:1}

  \item $\mu = \fact{s}{\atom a} \ \land\ c \abscmove{\atom a} c'
    \implies \realizes[\pmv A]{Q'}{c'}{s}$
    \label{eq:realizes-step:2}
  \end{enumerate}
\end{applemma}
\begin{proof}
  For part \ref{eq:realizes-step:1},
  let $\eta'$ be any $\setenum{\unblocked,\tell{}}$-fair do-free trace
  $Q'=P_0 \abspmove{} P_1 \abspmove{} \cdots$.
  Then, the trace $\eta$ defined as
  \[
  Q \abspmove{\mu} Q'=P_0 \abspmove{} P_1 \abspmove{} \cdots
  \]
  is a $\setenum{\unblocked,\tell{}}$-fair do-free trace of $Q$.

  To check Definition~\ref{def:realizes} on the trace $\eta'$,
  is suffices to exploit the same definition on $\eta$.
  Indeed, if Definition~\ref{def:realizes} applies to
  $\eta$, it also holds for $\eta'$ which is a suffix.

  For part \ref{eq:realizes-step:2}, consider any
  $\setenum{\unblocked,\tell{}}$-fair do-free trace of $Q$.  We have then
  $Q=P_0 \abspmove{\mu_0} \cdots$.  
  From Definition~\ref{def:realizes}, part \ref{def:realizes:2},
  taking $i=0,P'=Q'$, we get that
  \[
  Q=P_0 \abspmove{\fact{s}{\atom a}} Q' \land c \abscmove{\atom a} c'
  \implies
  \realizes{Q'}{c'}{s}
  \]
  which allows us to conclude.
\qed
\end{proof}

\begin{barttheorem}{Theorem~\ref{lem:realizes-step-co2}}
  \lemrealizesstepcoco
\end{barttheorem}
\begin{proof}
  By Lemma~\ref{lem:co2-to-abstract}, $(Q_i)_i$ forms an 
  abstract $\abspmove{}$-trace, whose labels $\mu_i$ are of
  the form $\pi$ if $S_i \pmove{\pmv A \says \pi} S_{i+1}$,
  and $\ctx$ otherwise.

  The compliance of $c_i$ and $d_i$ is preserved at each
  step by Def.\ref{def:compliance}. The fact that $Q_i$
  $\sharp$-realizes $c_i$ is also preserved by steps $Q_i
  \abspmove{\mu_i} Q_{i+1}$, as we now prove by cases on $\mu_i$.

  \begin{itemize}

  \item In the case $\mu_i = \ctx$, we get that $Q_{i+1}$ realizes $c_{i}$ by
    Lemma~\ref{lem:realizes-step}, item \ref{eq:realizes-step:1}. 
    If $c_{i+1} = c_i$ then $c_i \abscmove{\ctx} c_{i+1}$ trivially holds;
    otherwise, it has been modified by the context and
    by Lemma~\ref{lem:abstract-contract}, item 2,
    since $c_i \compliant d_i$ we again have $c_i \abscmove{\ctx} c_{i+1}$.
    From that, we apply Lemma~\ref{lem:realizes-ctx} to obtain that
    $Q_{i+1}$ also $\sharp$-realizes $c_{i+1}$.

  \item Otherwise, $\mu_i = \pi$, $S_i \pmove{\pmv A \says
      \pi} S_{i+1}$, and $Q_i \abspmove{\mu_i} Q_{i+1}$. We consider two
    further subcases.

    \begin{itemize}
    \item If $\mu_i = \fact{s}{\atom a}$, then 
      $S_i \pmove{\does[s]{\pmv A}{\atom a}} S_{i+1}$.
      The latter is due to a transition in contracts of the form
      $\pmv A \says c_i \mid \pmv B \says d_i
      \cmove{\pmv A \says \atom a}
      \pmv A \says c_{i+1} \mid \pmv B \says d_{i+1}$.
      So by Lemma~\ref{lem:abstract-contract}, item 1,
      we have that \mbox{$c_i \abscmove{\atom a} c_{i+1}$}. 
      Hence $Q_{i+1}$ $\sharp$-realizes $c_{i+1}$ by
      Lemma~\ref{lem:realizes-step}, item \ref{eq:realizes-step:2}.  

    \item If $\mu_i \neq \fact{s}{-}$, we have $c_{i+1} = c_i$ because
      $\pmv A$ did not perform any action in session $s$. Hence,
      we get that $Q_{i+1}$ $\sharp$-realizes $c_i = c_{i+1}$ from
      Lemma~\ref{lem:realizes-step}, item \ref{eq:realizes-step:1}.
      \qed

    \end{itemize}
  \end{itemize}
\end{proof}

\hidden{
\begin{definition}\label{def:taufairness}
  A trace $\eta = (P_i \abscmove \mu P_{i+1})_i$ is having length
  $|\eta| \in \mathbb{N} \cup \setenum{\infty}$ is
  $\setenum{\pi}$-fair if and only if, for all $i \in \mathbb{N}$
  such that $i \leq |\eta|$
  \[
  (\forall j \in [i , |\eta|]. \exists Q,R,P' \;. P_j = \pi.Q + R \mid
  P') \implies \exists h \geq i .\; \mu_h = \pi \land P_{h+1} = Q
  \mid P'
  \]
\end{definition}
}

\begin{applemma}\label{lem:par-abs}
  For all $P,Q$, $\pmv A$, $\mu$, if $P \mid Q \abspmove[\pmv A]\mu W $
  then $W \equiv P'
  \mid Q'$ and $\exists \sigma \;. P' \equiv P\sigma \vee \Big(P =
  \pi.P_1 + R_1 \mid P_2 \land P' \equiv (P_1 \mid P_2)\sigma \Big)$.

  Moreover, if $(P_i \mid Q_i)_i$ is a $\setenum{\unblocked,\tell{}}$-fair
  $\abspmove[]{}$-trace without $\fact{s}{-}$ where $P_0 = P$ and $Q_0
  = Q$, then  $(P_i)_i$ is a $\setenum{\unblocked,\tell{}}$-fair
  $\abspmove[]{}$-trace without $\fact{s}{-}$  from $P$.
\end{applemma}
\begin{proof}
  The proof of the first part easily follows by case analysis on the
  rules in Figure~\ref{fig:abstract-co2} observing that the abstract
  semantics does not allow parallel processes to interact.

  To prove that $(P_i)_i$ is a $\setenum{\unblocked,\tell{}}$-fair
  $\abspmove[]{}$-trace without $\fact{s}{-}$ from $P$ it suffices to
  note that any transition $P_i \mid Q_i \abspmove[\pmv A]\mu P_{i+1}
  \mid Q_{i+1}$ due to a prefix in $Q_i$ can be replaced with a
  transition $P_i \abspmove[\pmv A]\ctx P_{i+1}$ with a suitable
  substitution. Fairness then trivially holds.
  \qed
\end{proof}

\begin{applemma}\label{lem:realizes-xsafe}
  For all $P,Q$, $\pmv A$, $s$, $c$, if $\realizes[\pmv A]{P}{c}{s}$
  and $Q$ is free from $\fact{s}{-}$, then $\realizes[\pmv A]{P \mid
    Q}{c}{s}$.
\end{applemma}
\begin{proof}
  We prove that the relation
  \[
  \mathcal R = \setcomp{(P \mid Q \ ,\  c)}{ \realizes[\pmv A] P c s
    \land Q \text{ is free from } \fact s -}
  \]
  satisfies conditions~\ref{def:realizes:1} and~\ref{def:realizes:2}
  of Def.~\ref{def:realizes}.

  Let  $(P_i \mid Q_i)_i$ be a $\setenum{\unblocked,\tell{}}$-fair
  $\abspmove[]{}$-trace without $\fact{s}{-}$ where $P_0 = P$ and $Q_0
  = Q$.
  By Lemma~\ref{lem:par-abs},
  $(P_i)_i$ is a $\setenum{\unblocked,\tell{}}$-fair
  $\abspmove[]{}$-trace without $\fact{s}{-}$ from $P$.
  Since $Q_0 = Q$ is free from $\fact s -$, $\readydo s {Q_i} =
  \emptyset$ for each $i$; hence, $\readydo s {P_i \mid Q_i} =
  \readydo s {P_i}$.
  This, observing that $\realizes[\pmv A] P c s$, yields that there is
  an index $k$ such that $\unblocks{c}{\readydo{s}{P_j \mid Q_j}}$
  for each $j \geq k$.

  Finally, for each $i$, $\atom a$, $P'$, $c'$, if $P_i \mid Q_i
  \abspmove{\fact{s}{\atom{a}}} W$ and $c \abscmove{\atom a} c'$ then,
  by repeated application of Lemmata~\ref{lem:par-abs}
  and~\ref{lem:realizes-step}, $W$ is of the form $P' \mid Q_i$ with
  $Q_i$ without $\fact s -$ (hence $P_i \abspmove{\fact s {\atom a}}
  P'$) and $\realizes[\pmv A]{P'}{c'}{s}$.  Therefore, $(P' \mid Q_i,
  c') \in \mathcal R$ which implies that
  condition~\ref{def:realizes:2} of Def.~\ref{def:realizes} holds.
  \qed
\end{proof}

\begin{applemma}\label{lem:real-subs}
  For all $P$, $\pmv A$, $s$, $c$, if $\realizes [\pmv A] P c s$ and
  $\sigma$ is any substitution, then $\realizes [\pmv A] {P\sigma} c
  s$.
\end{applemma}
\begin{proof}
  Since $P \abspmove{\ctx} P\sigma$ for any substitution $\sigma$,
  the thesis is immediate from Lemma~\ref{lem:realizes-step}.
  \qed
\end{proof}

\begin{barttheorem}{Lemma~\ref{lem:canonical-step}}
  \lemcanonicalstep
\end{barttheorem}
\begin{proof}
  For part 1, let $\mathcal C$ be a $x$-safe context, and let $Q$ and
  $R$ be processes such that
  \begin{equation}
    P = \mathcal{C}(\tell{\pmv B}{\freeze{x}{c}}.Q + R)\label{eq:Pstr}
  \end{equation}
  By Def.~\ref{def:canonical}, either $\mathcal C(\bullet) = \mathcal
  C'((x)\bullet)$ or $\mathcal C$ does not contain $\fact{x}{-}$.
  The former case is ruled out by $P = \open{P}$.
  (Notice that there exists at least a context $\mathcal C$ not of the
  form $\pi. \mathcal C'(\bullet) + R' \mid Q'$ with $\pi \neq
  \tell{}{\freeze x c}$.)
  Therefore, there must be a context $\mathcal C = \bullet \mid Q'$
  such that~\eqref{eq:Pstr} holds.
  Hence, $P' = Q \mid Q'$ with $ \realizes [{\pmv A}]
  {Q\setenum{\bind{x}{s}}} c s $ and $Q'$ free
  from $\fact x {-}$ since $\pmv A[P]$ is \canonical\
  and~\eqref{eq:Pstr}; therefore the thesis follows by
  Lemma~\ref{lem:realizes-xsafe}.

  For part 2, by contradiction, assume $P'$ is not \canonical, then for suitable
  $\mathcal C(\bullet)$, $x$, $c$, $Q'$ and $R'$
  \begin{equation}
    \label{eq:not-canonical}
    P' = \mathcal C(\tell{\pmv B}{\freeze x c}.Q' + R')
    \quad \land \quad
    \Big(\notrealizes[{\pmv A}]{\open{Q'\setenum{\bind x s}}} c s
    \quad \vee \quad
    \mathcal C \text{ is not } x\text{-safe}\Big)
  \end{equation}
  We proceed by case analysis to derive a contradiction.
  \begin{itemize}
  \item
    If $P = \pi.Q + R_1 \mid P_2$, $\pi$ was fired, 
    and $\pi \neq \tell{\pmv A}
    {\freeze{x'}{c'}}$ then $P' = \open{Q \mid P_2}\sigma$ for some
    substitution $\sigma$.
    \newcommand{\tellPP}{\tell{\pmv B}{\freeze x c}.Q' + R'}
    Hence, $\tellPP$ is a sub-process of either of $\open{Q}\sigma$ or
    $\open{P_2}\sigma$.

    Note that, were $\sigma$ defined at $x$, we would have 
    that $\tell{\pmv B} {\freeze x c}.Q'$ is under a delimitation in
    $P'$, otherwise $P'$ could not contain $\tell{\pmv A}
    {\freeze x c}.Q'$ contradicting~\eqref{eq:not-canonical}.
    Hence, w.l.o.g. we can assume that $\sigma$ is not defined at $x$,
    otherwise we can $\alpha$-convert the bound variable.

    \newcommand{\tellP}{\tell{\pmv B}{\freeze x c}.Q'' + R''}
    Therefore there is a sub-process $\tellP$ of $P$ such that
    \[
    \open{\tellP}\sigma = \tell{\pmv B}{\freeze x c}.Q' + R'
    \]
    namely there is a context $\mathcal C_P(\bullet)$ such that $P =
    \mathcal C_P(\tellP)$, where $\mathcal C_P(\bullet)$ is $x$-safe
    by \canonicity\ of $\pmv A[P]$.
    %
    Observe that $\tellP$ cannot occur right under a delimitation of $x$,
    otherwise that would imply that $\mathcal C(\bullet)$
    is $x$-safe. This is impossible because we would have that
    $\notrealizes[{\pmv A}]{\open{Q'\setenum{\bind x s}}} c s$.
    However, by \canonicity\ of $\pmv A[P]$, we have
    $\realizes[\pmv A]{\open{Q''\setenum{\bind x s}}} c s$ hence
    \[
    \open{Q''\setenum{\bind x s}}\sigma = 
    \realizes[\pmv A]{\open{Q'\setenum{\bind x s}}} c s
    \]
    by Lemma~\ref{lem:real-subs} since $\setenum{\bind x s}\sigma =
    \sigma\setenum{\bind x s}$ because $x$ is not assigned by
    $\sigma$. This proves that $\tellP$ cannot occur right under
    a delimitation of $x$ in $\mathcal C_P$.
  
    Moreover, if $\fact x -$ does not occur in $\mathcal C_P(\bullet)$,
    then it can not occur in $\mathcal C(\bullet)$ as well, since
    transitions can not introduce it.

    From the above cases, we conclude that $\mathcal C_P$ is not $x$-safe,
    therefore $P$ is not \canonical --- contradiction.
    
  \item If $P = \pi.Q + R_1 \mid P_2$ and $\pi = \tell{\pmv A}
    {\freeze{x'}{c'}}$ then $P' = \open{\freeze{x'}{\pmv A \says c'} \mid Q
    \mid P_2}$ and the proof proceeds as in the previous case.

  \item If $P \abspmove{\ctx} \ \freeze{x'}{\pmv B \says c} \mid P$
    with $\pmv B \neq \pmv A$, then since no latent contracts 
    of the form $\freeze{y}{{\pmv A} \says d}$
    occur in ${\pmv A}[P]$, then this also holds for ${\pmv A}[P']$.
    By contradiction, were $\pmv A[\freeze{x'}{\pmv B \says c} \mid P]$ 
    non-\canonical\ then also $\pmv A[P]$ would be such. 
    This is because 
    $\mathcal C(\tell{\pmv A}{\freeze
      x c.Q' +R'}) =$ \mbox{$\freeze{x'}{\pmv B \says c} \mid P$} implies that
    $\mathcal C(\bullet) = \freeze{x'}{\pmv B \says c} \mid \mathcal
    C'(\bullet)$ with $P = \mathcal C'(\tell{\pmv A}{\freeze x c.Q'
      +R'})$.

  \item If $P \abspmove{\ctx} P \sigma$ and $P\sigma = \mathcal
    C(\tell{\pmv B}{\freeze x c.Q' +R'})$ then, as we did in the first
    case, w.l.o.g.~we can assume $\sigma$ to be undefined at
    $x$. Then, there is $\tellP$ and a context $\mathcal C_P(\bullet)$
    such that $(\tellP)\sigma = \tellPP$ and $P = \mathcal
    C_P(\tellP)$.
    Then the thesis is obtained  as in the first case.
\qed
  \end{itemize}
\end{proof}

\begin{applemma}\label{lem:solo-nonculpable}
If $\csmiley[s]{\pmv A}{S}$ and $S \pmove{\does[s]{\pmv A}{\atom a}} S'$
then $\csmiley[s]{\pmv A}{S'}$
\end{applemma}
\begin{proof}
By inspection of the semantics rules. In system $S$, consider 
the unilateral contract $c$ of $\pmv A$ in $s$. If $c$ performs
an $\atom e$ move, it changes into $\E$ and the thesis follows trivially.
Otherwise, $c$ can not start with $\ready{\atom a},
\atom a\neq \atom e$, since we have $\csmiley[s]{\pmv A}{S}$.
So, $c$ either moves according to a $\nrule{[*Fail]}$ rule,
or causes a $\ready{\atom a}, \atom a\neq \atom e$ to appear in front of
the other contract in session $s$. In both cases, we have
$\csmiley[s]{\pmv A}{S'}$.
\qed
\end{proof}

\pagebreak

\begin{barttheorem}{Theorem~\ref{th:canonical-is-saint}}
  \thcanonicalissaint
\end{barttheorem}
\begin{proof}
  By contradiction, assume that ${\pmv A}[P]$ is not honest.
  By Def.~\ref{def:saint}, there exists $S'$ 
  (free from contracts of the form ${\pmv A} \says c$)
  such that {\pmv A} is not honest in $S = \sys {\pmv A} P \mid S'$.
  By Def.~\ref{def:honest}, there exists a 
  $\pmove{}$-trace $S \pmove{}^* S_0$ and a
  $\setenum{\pmv A \says \cdots}$-fair {\pmv A}-solo 
  $\pmove{}$-trace $S_n,S_{n+1},\ldots$ 
  such that $\cfrown{\pmv A}{S_i}$ for all $i \geq n$.
  W.l.o.g.\ assume that such trace is stable.

  Note that, in the {\pmv A}-solo trace $S_n \pmove{} S_{n+1} \pmove{} \cdots$,
  sessions can only be initiated be the participant {\pmv A}.
  Since the environment of {\pmv A} in $S_n$ contains a \emph{finite}
  number of frozen contracts, and since sessions can only be established
  between two different participants, then a \emph{finite} number of sessions
  appears in $S_n \pmove{} S_{n+1} \pmove{} \cdots$.
  By Lemma~\ref{lem:solo-nonculpable}, for all sessions $s'$,
  if $\csmiley[s']{\pmv A}{S_i}$, then $\csmiley[s']{\pmv A}{S_{i+1}}$,
  i.e.\ {\pmv A} cannot become culpable by means of her own actions.
  Therefore, there exists a session $s$ and $n_s \geq n$ such that 
  $\cfrown[s]{\pmv A}{S_i}$ for all $i \geq n_s$.

  Therefore, $s$ contains a contract advertised by {\pmv A} at 
  some step $t < n$, which has been fused at some step $f$, for $t < f < n$,
  i.e.\ the trace has the form:
  \begin{align*}
    S_0 
    & \pmove{}^*  
    \pmove{{\pmv A} \says \tell{\pmv K}{\freeze{x}{c}}} 
    (\vec{u}_t) \ 
    \sys {\pmv A} {Q_t} \mid \sys {\pmv K} {\freeze{x}{{\pmv A} \says c} \mid \cdots} \mid S_t \\
    & \pmove{}^* \pmove{\;\;{\pmv K} \says \fuse{x}{}\;\;\;}
    (\vec{u}_f) \ 
    \sys {\pmv A} {Q_f} \mid \sys {\pmv K} {\cdots} \mid s[\bic{c_f}{d_f}] \mid S_f \\[4pt]
    & \pmove{}^* S_n \pmove{}^* \cdots
  \end{align*}
  where $c_f = c$. 
  By rule~\nrule{[Fuse]}, $c_f \compliant d_f$, and since compliance
  is preserved by $\pmove{}$-transitions, 
  $c_i \compliant d_i$ for all $i \geq f$.
  By Lemma~\ref{lem:co2-to-abstract}, there exists a
  $\setenum{\unblocked,\tell{}}$-fair $\abspmove[\pmv A]{}$-trace:
  \[
  Q_0
  \;\abspmove{\mu_0}\;
  Q_1
  \;\abspmove{\mu_1}\;
  \cdots
  \tag*{where $\mu_i = \pi$ iff $S_{i} \pmove{{\pmv A} \says \pi} S_{i+1}$}
  \]
  Note that $Q_i = \open{Q_i}$, because the trace is stable.
  Then, by Lemma~\ref{lem:canonical-step} (item~2),
  ${\pmv A}[Q_i]$ is \canonical\ for all $i$.
  By Lemma~\ref{lem:canonical-step} (item~1)
  $\realizes[\pmv A]{Q_t\setenum{\bind{x}{s}}}{c}{s}$.
  By Lemma~\ref{lem:realizes-step} (item 1), 
  for all $i \in [t,f-1]$, $\realizes[\pmv A]{Q_i\setenum{\bind{x}{s}}}{c}{s}$.
  By Theorem~\ref{lem:realizes-step-co2},
  for all $i \geq f$, $\realizes[\pmv A]{Q_i}{c_i}{s}$.
  Since $c_i \compliant d_i$ for all $i$, then by 
  Lemma~\ref{lem:compliant-fail} it follows that $c_i \neq \cnil$ for all $i$.
  Then, by Theorem~\ref{lem:doubledo} (used contrapositively),
  there exists $d \geq f$ such that $\mu_i \neq \fact{s}{-}$ for all $i \geq d$.
  Then, by Def.~\ref{def:realizes} (item 1),
  there exists $k \geq d$ such that 
  $\unblocks{c_i}{\readydo{s}{Q_i}}$
  for all $i \geq k$.
  Therefore by Lemma~\ref{lem:readydo} it follows that
  $\csmiley[s]{\pmv A}{S_k}$ or $S_k \pmove{{\pmv A} \says \fact{s}{\atom{a}}}$,
  but note that $\csmiley[s]{\pmv A}{S_k}$ is false by hypothesis.
  Hence, since the $\pmove{}$-trace is fair,
  then the prefix $\fact{s}{\atom{a}}$ should be eventually fired
  --- contradiction, because the trace no longer contains labels 
  ${\pmv A} \says \fact{s}{-}$ after the $d$-th step.
  \qed
\end{proof}


\end{document}